\documentclass[aps,twocolumn,prb,showpacs,floatfix]{revtex4}

\usepackage{epsfig,amsmath,amssymb}
\usepackage{graphics} 
\usepackage{subfigure}
\usepackage{color}
\usepackage{multirow}

\begin{document}

\title
{
Spin-1/2 $J_{1}$--$J_{2}$ Heisenberg model on a cross-striped square lattice
}
\author
{R.~F.~Bishop and P.~H.~Y.~Li}
\affiliation
{School of Physics and Astronomy, Schuster Building, The University of Manchester, Manchester, M13 9PL, UK}

\author
{C.~E.~Campbell}
\affiliation
{School of Physics and Astronomy, University of Minnesota, 116 Church Street SE, Minneapolis, Minnesota 55455, USA}

\begin{abstract}
  Using the coupled cluster method (CCM) we study the full
  (zero-temperature) ground-state (GS) phase diagram of a spin-half
  ($s=\frac{1}{2}$) $J_{1}$--$J_{2}$ Heisenberg model on a
  cross-striped square lattice.  Each site of the square lattice has 4
  nearest-neighbor exchange bonds of strength $J_{1}$ and 2
  next-nearest-neighbor (diagonal) bonds of strength $J_{2}$.  The
  $J_{2}$ bonds are arranged so that the basic square plaquettes 
  in alternating columns have either both or no $J_{2}$ bonds
  included.  The classical ($s \rightarrow \infty$) version of the
  model has 4 collinear phases when $J_{1}$ and $J_{2}$ can take
  either sign.  Three phases are antiferromagnetic (AFM), showing
  so-called N\'{e}el, double N\'{e}el and double columnar striped
  order respectively, while the fourth is ferromagnetic.  For the
  quantum $s=\frac{1}{2}$ model we use the 3 classical AFM phases as
  CCM reference states, on top of which the multispin-flip
  configurations arising from quantum fluctuations are incorporated in
  a systematic truncation hierarchy.  Calculations of the
  corresponding GS energy, magnetic order parameter and the
  susceptibilities of the states to various forms of valence-bond
  crystalline (VBC) order are thus carried out numerically to high
  orders of approximation and then extrapolated to the (exact)
  physical limit.  We find that the $s=\frac{1}{2}$ model has 5
  phases, which correspond to the four classical phases plus a new
  quantum phase with plaquette VBC order.  The positions of the 5
  quantum critical points are determined with high accuracy.  While
  all 4 phase transitions in the classical model are first order, we
  find strong evidence that 3 of the 5 quantum phase transitions in
  the $s=\frac{1}{2}$ model are of continuous deconfined type.
\end{abstract}


\maketitle

\section{INTRODUCTION}
\label{intro}
Magnetic models involving quantum spin systems on regular
two-dimensional (2D) lattices have been at the center of both
theoretical and experimental condensed matter research in recent times
(see, e.g., Refs.\ \onlinecite{2D_magnetism_1,2D_magnetism_2}).  Even
when such systems are described in terms of seemingly very simple
Hamiltonians they can often display a bewildering variety of
ground-state (GS) phases with different types of ordering, even at
zero temperature ($T=0$).  The phases of the quantum systems (with
spins of a finite nonzero value of the spin quantum number, $s$), and
the transitions between them as some internal control parameter is
varied across the corresponding quantum critical points (QCPs), often
differ widely from those of their classical ($s \rightarrow \infty$)
counterparts.  Such control parameters usually provide a measure of
the degree of dynamic frustration between competing interactions in the
system.  Since it is widely believed that many of the properties of a
large variety of interesting strongly-correlated quantum many-body
systems can be understood in terms of the competition between GS
phases with qualitatively different properties, particular interest
has focussed on the associated quantum phase transitions and the
behavior of the system near such QCPs (see, e.g., Refs.\
\onlinecite{Sachdev:1999,Sentil:2004,Sachdev:2011}).

The often subtle interplay between frustration and quantum
fluctuations can lead to quantum spin-lattice models exhibiting $T=0$
GS phase diagrams that are very different from their classical
counterparts.  Since quantum effects tend to diminish as the spin
quantum number increases, spin-1/2 models have a special role to play.
Whereas many phase transitions in classical systems are often of
first-order type, where the transition between the two phases involves
sudden jumps in many of the physical properties, quantum fluctuations
can even act to turn such a first-order classical transition into a
(continuous) second-order one.  Precisely at or very near such a
second-order QCP the GS phase has very special properties.  The GS
wave function becomes a complex superposition of an exponentially
large set of multispin configurations that fluctuate at all length
scales, and hence it exhibits long-range entanglement phenomena.  Such
GS wave functions are very different from those of the quasiclassical
states that often lie on one or other (or both) sides of the QCP in
the GS phase diagram, and whose wave functions can be described in
simple terms as a product state for all the spins.  Of course, quantum
fluctuations still produce a mixture of other ``wrong'' multispin
configurations on top of such a simple product state, but far from the
QCP they do not totally destroy the classical order present in the
simple quasiclassical state.

The traditional Landau-Ginzburg-Wilson (LGW)\cite{Landau:1999,Wilson:1974}
description of quantum second-order transitions and their associated
critical singularities and quantum critical phenomena has been
remarkably successful in describing many quantum phase transitions.
However, it has become clear in recent years that there are other
continuous transitions that do not fit the LGW paradigm for critical
phenomena in which the critical singularities are associated with the
fluctuations of some appropriate order parameter that captures the
essential difference between the two phases on either side of the
transition.  In particular, it has been shown\cite{Sentil:2004}
how subtle quantum interference effects can invalidate the LGW
paradigm of QCPs separating phases characterized by such standard {\it
  confining} order parameters, by the appearance of an emergent gauge
field and consequent {\it deconfined} degrees of freedom which are
associated with the fractionalization of the appropriate order
parameters.  Thus, in this alternate deconfined scenario an emergent
gauge field mediates the interactions between the associated emergent
particles that carry fractions of the quantum numbers corresponding to
the underlying degrees of freedom.  Such fractional particles are
confined at low energies so that they do not appear sufficiently far
away on either side of the QCP, but they emerge naturally (and hence
deconfine) just at the QCP.  Such deconfined second-order phase
transitions can occur between states that break different symmetries, a
scenario which is not allowed in the standard LGW description.

Since quantum-critical states themselves are so inherently complex
they have largely been studied using either techniques from quantum
field theory or large-scale numerical simulations, usually of the
quantum Monte Carlo (QMC) kind.  It is clear that near QCPs,
where the effects of quantum fluctuations, and hence the complexity of
the wave functions, increase the closer one approaches them, very accurate
quantum many-body techniques are needed.  One such method is the
coupled cluster method (CCM),\cite{Bi:1991,Bishop:1998,Fa:2004}
which has very successfully been applied to a wide variety of
frustrated quantum spin-lattice models (see, e.g., Refs.\
\onlinecite{Fa:2004,Ze:1998,Bishop:1998_J1J2mod,Kr:2000,Bishop:2000,Fa:2001,Darradi:2005,Schm:2006,Bi:2008_PRB_J1xxzJ2xxz,Bi:2008_JPCM,darradi08,Bi:2009_SqTriangle,Darradi:2009_J1J2_XXZmod,richter10:J1J2mod_FM,Bishop:2010_UJack,Bishop:2010_KagomeSq,Reuther:2011_J1J2J3mod,DJJF:2011_honeycomb,Gotze:2011,Bishop:2012_honey_phase,Bishop:2012_checkerboard,Li:2012_honey_full,Bishop:2012_honeyJ1-J2,Li:2012_anisotropic_kagomeSq,RFB:2013_hcomb_SDVBC,Li:2013_chevron}),
yielding a good description of their $T=0$ GS phase diagrams and
accurate numerical values for their QCPs, even ones involving
deconfined transitions (see, e.g., Refs.\
\onlinecite{DJJF:2011_honeycomb,Bishop:2012_honeyJ1-J2,Li:2012_honey_full,RFB:2013_hcomb_SDVBC}).

The CCM has been found to yield accurate descriptions of quantum
magnets with different types of both quasiclassical magnetic order and
quantum paramagnetic order, including various types of valence-bond
crystalline (VBC) order.  These include the prototypical
$J_{1}$--$J_{2}$ model on the square lattice,
\cite{Bishop:1998_J1J2mod,darradi08} which contains nearest-neighbor
(NN) Heisenberg interactions with exchange coupling strength $J_{1}$
and corresponding next-nearest-neighbor (NNN) (i.e., diagonal) bonds
of strength $J_{2}$, as well as models that generalize it by
introducing both spatial lattice anisotropy\cite{Bi:2008_JPCM} and
spin anisotropy.\cite{Bi:2008_PRB_J1xxzJ2xxz,Darradi:2009_J1J2_XXZmod}  Of particular
interest for present purposes, they also include models in the
so-called half-depleted $J_{1}$--$J_{2}$ class, in the sense that they
are obtained from the full $J_{1}$--$J_{2}$ model on the square
lattice by removing half of the $J_{2}$ bonds in different
arrangements.  Examples include: (a) the ($J_{1}$--$J_{2}'$ or)
interpolating square-triangle model,\cite{Bi:2009_SqTriangle} (b) the
so-called Union Jack model,\cite{Bishop:2010_UJack} (c) the
anisotropic planar pyrochlore (APP) model (also known as the crossed chain
model) that comprises a $J_{1}$--$J_{2}$ model on the checkerboard
lattice,\cite{Bishop:2012_checkerboard} and (d) an analogous
$J_{1}$--$J_{2}$ model on a chevron-square lattice.\cite{Li:2013_chevron}

This half-depleted $J_{1}$--$J_{2}$ class of models on the square
lattice exhibits a wide variety of GS phase diagrams and QCPs, all of
which pose serious theoretical challenges, but which serve to improve
our understanding of quantum critical phenomena.  In the current paper
we consider another member of this class, namely the spin-1/2
$J_{1}$--$J_{2}$ Heisenberg model on a cross-striped square lattice,
and we find that it too has a rich $T=0$ GS phase diagram that
includes many of the features discussed above.

As a motivation for this study we note that while it is true that
there are many different ways that a Heisenberg antiferromagnet might
be frustrated in principle, the $J_{1}$--$J_{2}$ model and its
depleted counterparts occupy a central role.  Of the half-depleted
class described above there are two principal sub-classes.  The first
is where each square plaquette (formed from four NN $J_{1}$ bonds)
contains one $J_{2}$ bond.  The three main members of this sub-class
have been well studied previously.  They comprise (a) the
interpolating square-triangle lattice model,\cite{Bi:2009_SqTriangle}
(b) the Union Jack lattice model,\cite{Bishop:2010_UJack} and (c) the
chevron-decorated square lattice model.\cite{Li:2013_chevron} These
have the respective features that the arrangements of the remaining
diagonal NNN $J_{2}$ bonds are such that in case (a) the $J_{2}$ bonds
are all parallel, while in case (b) the orientations of the $J_{2}$
bonds alternate along both rows and columns, and in case (c) the
orientations of the $J_{2}$ bonds is the same along one square-lattice
axis direction (say, rows) but alternates along the perpendicular
direction (say, columns).  The second principal sub-class is where
half the basic square plaquettes have both $J_{2}$ bonds (namely, the
filled squares) while the remainder have neither (namely, the empty
squares).  There are clearly now two main members of this sub-class,
namely (i) where the empty and filled squares alternate along both rows
and columns (which is precisely the anisotropic planar pyrochlore
model or checkerboard lattice model\cite{Bishop:2012_checkerboard}
that has received much previous attention), and (ii) where alternating
columns, say, comprise filled squares and empty squares (which is
precisely the present cross-striped square lattice model).  From among
all members of the above two sub-classes only the latter model seems
not to have been studied before now.

\begin{figure*}[t]
\begin{center}
\vspace{0.7cm}
\hspace{-0.3cm}
\mbox{
\subfigure[]{\scalebox{0.4}{\includegraphics{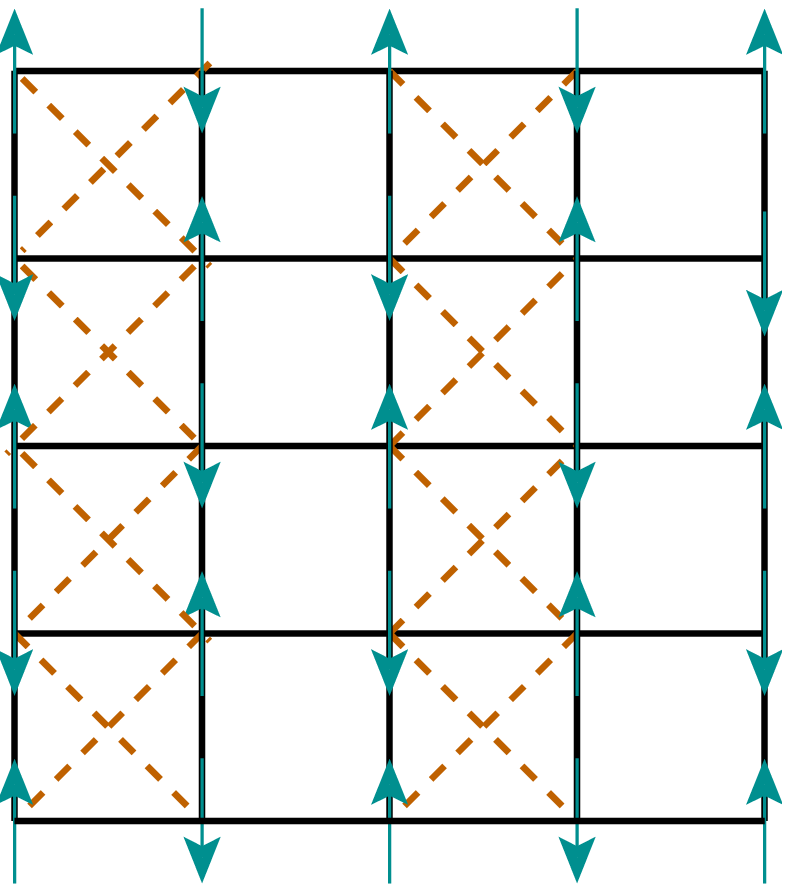}}} \hspace{0.05cm}
\subfigure[]{\scalebox{0.4}{\includegraphics{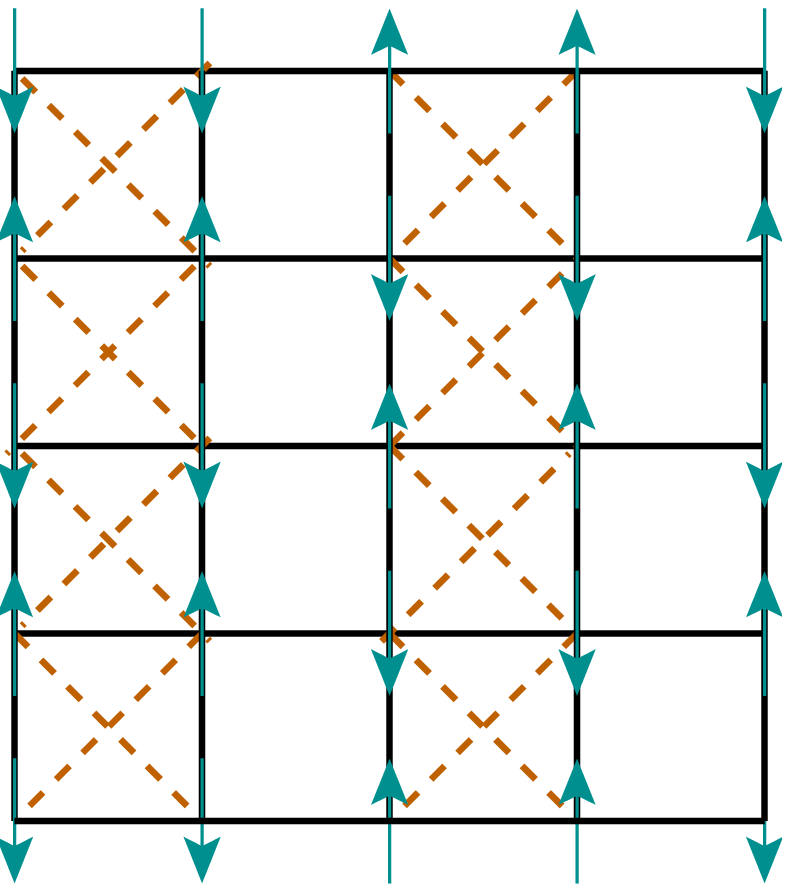}}}  \hspace{0.05cm}
\subfigure[]{\scalebox{0.4}{\includegraphics{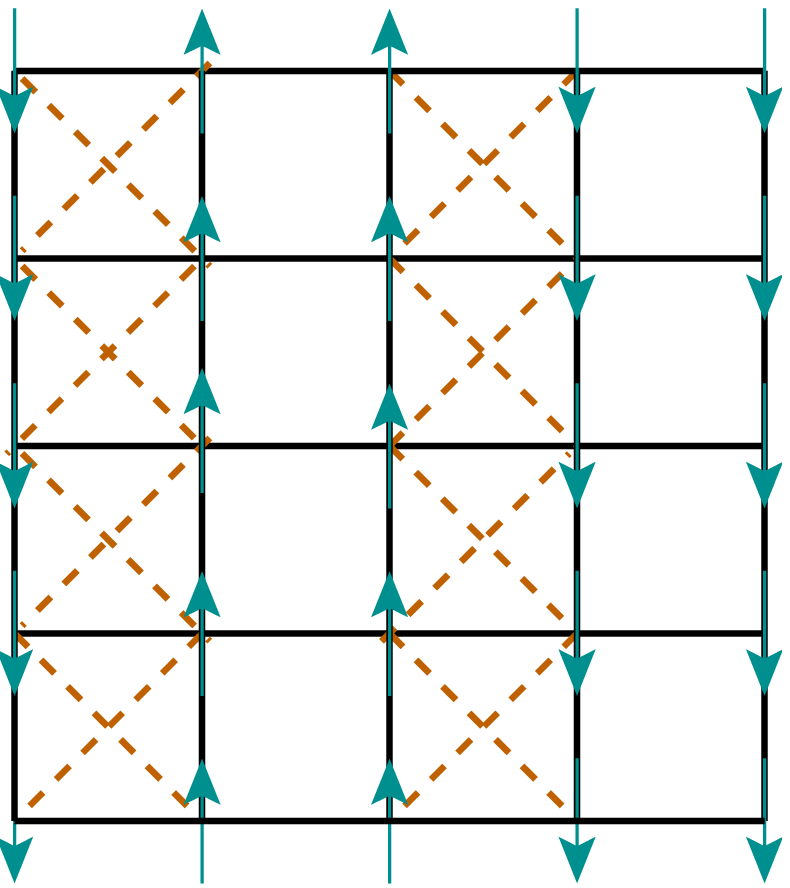}}}
\hspace{1.7cm}\subfigure[]{\scalebox{0.35}{\includegraphics[angle=90]{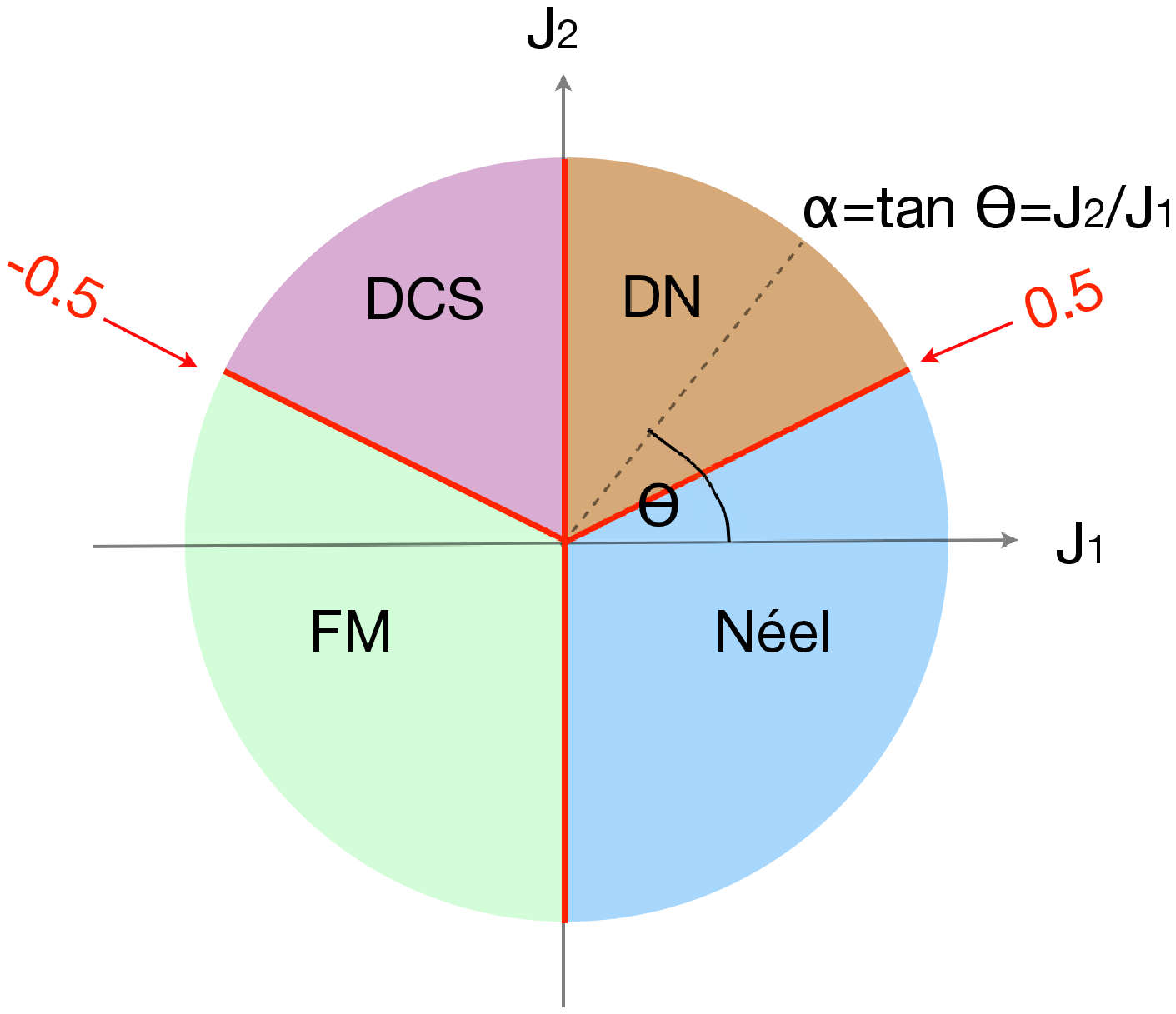}}} 
}
\end{center}
\caption{(Color online) The $J_{1}$--$J_{2}$ Heisenberg model on a cross-striped
  square lattice.  The solid (black) lines are $J_{1}$ bonds and the
  dashed (brown) lines are $J_{2}$ bonds.  The (cyan) arrows represent the
  relative spin directions in: (a) the N\'{e}el state, (b) the double
  N\'{e}el (DN) state, and (c) the double columnar striped (DCS)
  state.  (d) The classical phase diagram of the model.}
\label{model_bonds}
\end{figure*}

As additional motivation for this study we note that the successful
continual syntheses of new quasi-2D magnetic materials, and the
experimental observations of their properties, provides an ongoing
challenge for the theorist.  While we are unaware of any experimental
realization of the current cross-striped square-lattice model, several
already exist for other members of the depleted $J_{1}$--$J_{2}$
class.  For example, for the related interpolating square-triangle model,
\cite{Bi:2009_SqTriangle} the magnetic material Cs$_{2}$CuCl$_{4}$
provides a good experimental realization.  We suspect that similar
candidates will soon emerge as realizations of the current model.
Perhaps even more exciting, however, is the prospect of being able to
realize spin-lattice models with ultracold atoms trapped in
appropriate optical lattices,\cite{Struck:2011} with the subsequent
ability to tune the strengths of the competing NN $J_{1}$ bonds and
NNN $J_{2}$ bonds, and hence to drive the system from one GS phase to
another, thereby exploring the QCPs experimentally.

We now describe the model in detail in Sec.\ \ref{model_sec}, before
outlining the main features of the CCM in Sec.\ \ref{ccm_sec}.  Our
results are presented and discussed in Sec.\ \ref{results_sec}, and
we summarize and conclude in Sec.\ \ref{summary_sec} where the full
GS phase diagram for the model is presented.

\section{THE MODEL}
\label{model_sec}
The model considered here is the so-called $J_{1}$--$J_{2}$ Heisenberg model on
the cross-striped square lattice.  Its Hamiltonian is written as
\begin{equation}
H = J_{1}\sum_{\langle i,j \rangle} \mathbf{s}_{i}\cdot\mathbf{s}_{j} + J_{2}\sum_{\langle\langle i,k \rangle\rangle'} 
\mathbf{s}_{i}\cdot\mathbf{s}_{k}\,, \label{H}
\end{equation}
where the operators
$\mathbf{s}_{i}\equiv(s^{x}_{i},s^{y}_{i},s^{z}_{i}$) are the usual
quantum spin operators on lattice site $i$, with ${\bf s}^{2}_{i}=s(s+1)$.
Here we concern ourselves only with the extreme quantum case where all
lattice sites are occupied by a spin with spin quantum number
$s=\frac{1}{2}$.  On the underlying square lattice the sum over
$\langle i,j \rangle$ in Eq.\ (\ref{H}) runs over all distinct NN bonds
(each of which has the same exchange coupling strength $J_{1}$),
whereas the corresponding sum over $\langle\langle i,k
\rangle\rangle'$ runs over only half of the distinct NNN (diagonal)
bonds (each of which has the same exchange coupling strength $J_{2}$).
In the latter sum half of the basic square plaquettes, the so-called
filled squares, have both diagonal ($J_{2}$) bonds included, while the
remainder, the empty squares, have neither diagonal bond included.  The
pattern of the filled and empty squares, as shown in
Fig.\ \ref{model_bonds}, is such that along one of the basic
square-lattice directions (say, along rows) filled and empty squares
alternate, while along the perpendicular direction (say, along
columns) the squares are either all empty or all filled.  
Thus, the $J_{1}$--$J_{2}$ model on the cross-striped square lattice differs
from the corresponding APP model on the checkerboard lattice simply by
the pattern of filled and empty squares.  Both models contain equal
numbers of filled and empty squares, but in the APP model the empty and
filled squares alternate along both rows and columns.  The primitive
unit cell on the cross-striped square lattice, as shown in
Fig.\ \ref{model_bonds}, thus has size $2 \times 1$.  In both sums in
Eq.\ (\ref{H}) each bond is counted only once.

We are interested here in the full $T=0$ GS phase diagram of the
model, and hence in each of the cases where both types of bonds are
(independently) either ferromagnetic (FM) or antiferromagnetic (AFM)
in nature.  Since the overall energy scale is irrelevant for the phase
diagram, once we have specified the sign of either $J_{1}$ or $J_{2}$,
the model is completely specified by the ratio $\alpha \equiv
J_{2}/J_{1}$.  Let us first consider the simpler case when $J_{2}<0$
(and hence the NNN bond is FM in character).  In this case the
(independent) one-dimensional (1D) zigzag chains joined by $J_{2}$
bonds prefer to have FM order, and for either sign of $J_{1}$ the
system is unfrustrated since the ordering directions of different
chains is not fixed by the sign of $J_{2}$ alone.  Thus, if $J_{1}<0$
the classical ($s \rightarrow \infty$) system will take overall FM
ordering, while for $J_{1}>0$ the system will take N\'{e}el AFM
ordering.  Conversely, in the more complicated case when $J_{2}>0$,
such that the 1D zigzag chains connected by $J_{2}$ bonds prefer
N\'{e}el AFM order along them, the $J_{1}$ bonds will act to frustrate
this order for either sign of $J_{1}$.

In order to place our later work for the $s=\frac{1}{2}$ model in
context, let us first consider its classical ($s \rightarrow \infty$)
counterpart.  It is straightforward to show that the classical
$J_{1}$--$J_{2}$ model on the cross-striped square lattice has four GS
phases, separated by four first-order phase transitions, all as shown
in Fig.\ \ref{model_bonds}(d).  Firstly, the N\'{e}el AFM phase, which
is shown in Fig.\ \ref{model_bonds}(a), and which forms the stable GS
phase for $J_{1}>0$ and $J_{2}=0$, persists as the frustration is
increased (i.e., with $J_{2} > 0$) until a first critical point
$\alpha^{{\rm cl}}_{1} \equiv \frac{1}{2}$, where it undergoes a
first-order phase transition to another collinear AFM phase, the
so-called double N\'e{e}l (DN) phase, shown in Fig.\
\ref{model_bonds}(b).  This state has AFM N\'{e}el ordering along the
square-lattice axis direction parallel to the cross-stripes of filled
squares (i.e., along columns in Fig.\ \ref{model_bonds}), but with
spins alternating in a pairwise fashion in the perpendicular direction
(i.e., along rows in Fig.\ \ref{model_bonds}), such that on filled
(empty) squares NN spins are parallel (antiparallel) in the row
direction.  This DN state itself now persists as the frustration
parameter $\alpha$ is further increased.  If we define $\alpha \equiv
\tan \theta$, such that the full phase diagram is specified by
$\theta$ in the range $0 \leq \theta < 2\pi$, then the DN state is the stable GS phase in
the range $\theta^{{\rm cl}}_{1}<\theta<\theta^{{\rm cl}}_{2}$, with
$\theta^{{\rm cl}}_{1}=\tan^{-1}(\frac{1}{2})$ and $\theta^{{\rm
    cl}}_{2}=\frac{1}{2}\pi$.  At the second critical point,
$\theta^{{\rm cl}}_{2}$, as $J_{1}$ now is allowed to become negative
(i.e., so that $\theta$ increases beyond $\frac{1}{2}\pi$), the DN
phase gives way to a third collinear AFM phase, the so-called double
columnar striped (DCS) phase shown in Fig.\ \ref{model_bonds}(c).  Both
the DN and DCS phases have N\'{e}el ordering along the independent 1D
zigzag chains joined by $J_{2}$ bonds.  Like the DN state, the DCS
state also has spins alternating in a pairwise fashion in the
square-lattice axis direction perpendicular to the cross-stripes of
filled square (i.e., along rows in Fig.\ \ref{model_bonds}), but now
such that on empty (filled) squares NN spins are parallel
(antiparallel) in the row direction.  On the other hand, in the
perpendicular direction (i.e., along columns in Fig.\
\ref{model_bonds}), the spins in the DCS state form FM chains, with
the orientation of the spins now alternating in a pairwise fashion, as
shown in Fig.\ \ref{model_bonds}(c).  This DCS state itself forms the
stable GS phase over the range $\theta^{{\rm
    cl}}_{2}<\theta<\theta^{{\rm cl}}_{3}$, with $\theta^{{\rm
    cl}}_{3}=\tan^{-1}(-\frac{1}{2})$.  Finally, at $\alpha^{{\rm
    cl}}_{3}=-\frac{1}{2}$ the DN phase gives way to the FM phase, which
itself persists over the range $\theta^{{\rm
    cl}}_{3}<\theta<\theta^{{\rm cl}}_{4}$, with $\theta^{{\rm
    cl}}_{4}=\frac{3}{2}\pi$.  Finally, at the fourth critical point,
$\alpha^{{\rm cl}}_{4}$, there is a first-order transition between the
FM and N\'{e}el AFM phases.

Compared to the classical ($s \rightarrow \infty$) version of the
$J_{1}$--$J_{2}$ model on the cross-striped square lattice, the GS
phase of the $s=\frac{1}{2}$ model is really only well established at
a few special values of the parameter $\theta$.  Firstly, for the case
$\theta=0$, corresponding to the isotropic square-lattice Heisenberg
antiferromagnet (HAF), essentially all methods now concur that the
classical N\'{e}el AFM long-range order (LRO) is not destroyed.
Nevertheless, the staggered magnetization is reduced from its
classical value of 0.5 by quantum fluctuations, and the basic
excitations are gapless magnons with integer spin values.  Similarly,
at the point $\theta=\frac{1}{2}\pi$ in the phase diagram, of the
$s=\frac{1}{2}$ model, we have the well-known and exactly soluble case
of uncoupled 1D HAF chains.  Such 1D spin-1/2 chains have a Luttinger
spin-liquid GS phase, on top of which there exists a gapless
excitation spectrum of deconfined spin-1/2 spinons.

Apart from the above two points and the obvious regime $\pi \leq
\theta \leq \frac{3}{2}\pi$ (i.e., where $J_{1} \leq 0$ and $J_{2}
\leq 0$) where the FM state, which is always an exact eigenstate of
the Hamiltonian of Eq.\ (\ref{H}) for any value of the spin quantum
number $s$, provides the actual GS phase, little else is known with
any precision about the GS phase diagram for the spin-1/2 case.
Nevertheless, various plausible conjectures may be made.  For example,
one expects from continuity that the (partial) N\'{e}el order present
at $\theta = 0$ should survive as the frustrating $J_{2}$ bonds (i.e.,
with $J_{2}>0$) are slowly turned on and increased in strength, all
the way out to some critical value, at which the
N\'{e}el order (i.e., the staggered magnetization) goes to zero.  It
also seems plausible that as one moves away from the $\theta = 0$
point in the opposite direction (i.e., with $J_{2} <0$), the FM
$J_{2}$ bonds should now strengthen the N\'{e}el order.  Thus, one has
no {\it a priori} reason to expect that the lower half of the phase
diagram (i.e., with $J_{2}<0$) in Fig.\ \ref{model_bonds} should
differ between the classical ($s \rightarrow \infty$) and the quantum
($s=\frac{1}{2}$) cases.

Much more tentatively, one might be tempted to expect that in the
large-$\alpha$ region near $\theta=\frac{1}{2}\pi$ the 1D Luttinger
behavior, which is present precisely at this $J_{1}=0$ limit, might
also be robust against the turning on of the interchain ($J_{1}$)
couplings, so that the spin-1/2 chains effectively continue to act as
decoupled.  Such a 2D quantum spin liquid (QSL) GS phase would provide
an example of what has been termed a sliding Luttinger liquid
(SLL).\cite{Emery:2000, Mukhopadhyay:2001,Vishwanath:2001}

Such an SLL phase was predicted to occur\cite{Starykh:2002} in the related
spin-1/2 APP model on the checkerboard lattice, which we have
mentioned previously.  Nevertheless, a later more detailed
study\cite{Starykh:2005} of the relevant terms near the 1D
Luttinger liquid fixed point showed that this earlier prediction of an
SLL phase was erroneous.  In the same
analysis\cite{Starykh:2005} it was suggested that the correct GS
phase in this limiting regime might instead exhibit a form of
VBC order, in which the system dimerizes
with a staggered ordering of dimers along the corresponding $J_{2}$
chains of the APP model.  In an
analysis\cite{Bishop:2012_checkerboard} of the spin-1/2
$J_{1}$--$J_{2}$ (APP) model on the checkerboard lattice, using the
same methodology as we apply here to the spin-1/2 $J_{1}$--$J_{2}$
model on the cross-striped square lattice, firm evidence was found for
this so-called crossed-dimer VBC (CDVBC) GS phase for all values of the
ratio $J_{2}/J_{1}$ (with $J_{1}>0$) above an upper critical value.
Clearly it will be of considerable interest to investigate, as part of
the present study, what is the GS phase of the present model in this
same very interesting and most challenging regime.

In Sec.\ \ref{ccm_sec} below we first outline the most salient and
most important features of the coupled cluster method (CCM) that we
use here, before discussing our results obtained from it in Sec.\ \ref{results_sec}.  We then end in Sec.\ \ref{summary_sec} with a brief
summary and conclusions.

\section{THE COUPLED CLUSTER METHOD}
\label{ccm_sec}
The coupled cluster method (CCM) has become one of the most pervasive
and most accurate (at attainable levels of computational implementation)
of all modern techniques of quantum many-body theory (see, e.g.,
Refs.\
\onlinecite{Bi:1991,Bishop:1998,Fa:2004,Bishop:1987,Arponen:1991} and references cited therein).  It has
been applied very successfully to a wide variety of quantum many-body
systems in many fields, including quantum chemistry, atomic and
molecular physics, condensed matter physics, nuclear physics and
subnuclear physics.  Of particular interest for present purposes is
its wide usage in recent years to investigate the GS phase structure of a
large number of spin-lattice models of interest in quantum magnetism
(see, e.g., Refs.\
\onlinecite{Fa:2004,Ze:1998,Bishop:1998_J1J2mod,Kr:2000,Bishop:2000,Fa:2001,Darradi:2005,Schm:2006,Bi:2008_PRB_J1xxzJ2xxz,Bi:2008_JPCM,darradi08,Bi:2009_SqTriangle,Darradi:2009_J1J2_XXZmod,richter10:J1J2mod_FM,Bishop:2010_UJack,Bishop:2010_KagomeSq,Reuther:2011_J1J2J3mod,DJJF:2011_honeycomb,Gotze:2011,Bishop:2012_honey_phase,Bishop:2012_checkerboard,Li:2012_honey_full,Bishop:2012_honeyJ1-J2,Li:2012_anisotropic_kagomeSq,RFB:2013_hcomb_SDVBC,Li:2013_chevron}
and references cited therein).  The CCM provides a systematic means to
investigate various possible GS phases and their regions of stability,
including an accurate determination of the associated QCPs.  The
description is, in every case, capable of systematic improvement in
accuracy, since it is formulated in terms of well-defined hierarchical
approximation schemes, which incorporate an increasing number of the
multispin-flip configurations that are present in the exact GS quantum
many-body wave function as the level of truncation is improved.

The CCM formalism is well described in the literature (see, e.g.,
Refs.\
\onlinecite{Bi:1991,Bishop:1998,Ze:1998,Kr:2000,Bishop:2000,Fa:2004,Bishop:1987,Arponen:1991}
and references cited therein), and hence we only outline briefly its
key ingredients as required for the present study.  To implement the
CCM one always needs to choose a so-called (normalized) model or
reference state $|\Phi\rangle$.  This is often conveniently (but not
necessarily) chosen as a classical state, which may or may not form an
actual GS phase of the classical counterpart of the model (i.e., its $s
\rightarrow \infty$ counterpart for spin-lattice models) in some
region of the $T=0$ GS phase diagram parameter space.  For our present
study of the $J_{1}$--$J_{2}$ model on the cross-striped square
lattice we will present results in Sec.\ \ref{results_sec} below
based in turn on each of the three classical states shown in Figs.\
\ref{model_bonds}(a)--(c) as CCM model states.

The exact, fully correlated, GS ket- and bra-state wave functions of the interacting system are denoted as 
$|\Psi\rangle$ and $\langle\tilde{\Psi}|$ respectively, with normalizations chosen to satisfy $\langle\tilde{\Psi}|\Psi\rangle = \langle{\Phi}|\Psi\rangle =
\langle{\Phi}|\Phi\rangle = 1$.  They are now
parametrized in terms of the CCM reference state as
\begin{equation}
|\Psi\rangle={\rm e}^{S}|\Phi\rangle\,; \qquad \langle\tilde{\Psi}|=\langle\Phi|\tilde{S}{\rm e}^{-S}\,, \label{ket_bra_eq}
\end{equation}  
where the exponential forms lie at the heart of the method.  The ket- and bra-state correlation operators, $S$ and $\tilde{S}$ respectively, now incorporate explicitly the multispin-flip configurations in $|\Psi\rangle$ and $\langle\tilde{\Psi}|$ beyond those contained in the chosen model state $|\Phi\rangle$, caused by quantum fluctuations.  Hence, they are expanded as
\begin{equation}
  S=\sum_{I \neq 0}{\cal S}_{I}C^{+}_{I}\,; \qquad \tilde{S}=1+\sum_{I \neq 0}\tilde{{\cal S}}_{I}C^{-}_{I}\,,  \label{S_operators}
\end{equation}
where we define $C^{+}_{0} \equiv 1$ to be the identity operator and
where the set-index $I$ represents a particular set of lattice spins.
It is used to encode any particular multispin-flip configuration with
respect to state $|\Phi\rangle$, such that $C^{+}_{I}|\Phi\rangle$ is
the corresponding wave function for this configuration.  Thus the
operator $C^{+}_{I} (\equiv (C^{-}_{I})^{\dagger})$ may be regarded as
a multispin-flip creation operator with respect to $|\Phi\rangle$,
which itself acts as a generalized vacuum state.  It is important to
note that these operators must also be chosen to satisfy the relations
$C^{-}_{I}|\Phi\rangle=0=\langle\Phi|C^{+}_{I}$, which reinforce their
interpretation as given above.  The choice of the set-indices $\{I\}$
and the operators $\{C^{+}_{I}\}$ is discussed more fully below.

The subsequent implementation of the CCM for spin-lattice systems is
considerably simplified if one now chooses a set of local coordinate
frames in spin space, which must be chosen separately for each model
state used, such that on each lattice site in each model state the
spin aligns in the downward (i.e., along the negative $z$ axis)
direction.  Such passive rotations clearly leave the basic SU(2) spin
commutation relations unchanged, and hence cause no physical effects.
However, this simple choice has the consequence that in this basis the
$C^{+}_{I}$ operators take the universal form, $C^{+}_{I} \equiv
s^{+}_{j_{1}}s^{+}_{j_{2}}\cdots s^{+}_{j_{n}}$, of being products of
single-spin raising operators, $s^{+}_{j} \equiv
s^{x}_{j}+is^{y}_{j}$, and the set-index $I \equiv
\{j_{1},j_{2},\cdots , j_{n};\; n=1,2,\cdots , N\}$ is simply a set
of lattice site indices, with $N$ being the total number of sites.
Clearly, for a spin with spin quantum number $s$, the raising operator
$s^{+}_{j}$ on a given site $j$ may be applied a maximum number of $2s$
times, and hence a given site-index may appear a maximum of 2$s$
times in any set-index $I$ included in the sums in Eq.\ 
(\ref{S_operators}).  Hence for the present $s=\frac{1}{2}$ case, no
single site-index $j_{k}$ may appear more than once in any set-index
$I$.

The CCM thus encapsulates the correlations present in the exact GS
phase in terms of the ket- and bra-state correlation coefficients
$\{{\cal S}_{I},{\tilde{\cal S}}_{I}\}$, and these may now formally be
calculated by minimization of the the GS energy expectation functional,
$\bar{H}=\bar{H}({\cal S}_{I},{\tilde{\cal S}_{I}})\equiv
\langle\tilde{\Psi}|H|\Psi\rangle$, where $H$ is the Hamiltonian of
the system, with respect to each of the coefficients $\tilde{\cal S}_{I}$
and ${\cal S}_{I}$, $\forall I \neq 0$.  A simple use of Eqs.\
(\ref{ket_bra_eq}) and (\ref{S_operators}) then leads respectively to
the coupled sets of equations $\langle\Phi|C^{-}_{I}{\rm e}^{-S}H{\rm
  e}^{S}|\Phi\rangle=0$ and $\langle\Phi|\tilde{S}{\rm
  e}^{-S}[H,C^{+}_{I}]{\rm e}^{S}|\Phi\rangle = 0, \forall I \neq 0$.
Clearly, these equation are completely equivalent to the GS
ket- and bra-state Schr\"{o}dinger equations, $H|\Psi\rangle \equiv
E|\Psi\rangle$ and $\langle\tilde{\Psi}|H \equiv
E\langle\tilde{\Psi}|$.  The CCM equations for the bra-state
correlation coefficients may be written equivalently in the form
$\langle\Phi|\tilde{S}({\rm e}^{-S}H{\rm
  e}^{S}-E)C^{+}_{I}|\Phi\rangle = 0, \forall I \neq 0$.

Clearly, the CCM ket-state equations for the set of $c$-number
correlation coefficients $\{{\cal S}_{I}\}$ are intrinsically
nonlinear, due to the presence of the operator $S$ in Eq.\ 
(\ref{ket_bra_eq}) in the exponentiated form e$^{S}$.
Nevertheless, it is another key feature of the CCM that in the
equations we actually solve for the correlation coefficients it only
ever appears in the form of the similarity transform of the
Hamiltonian, ${\rm e}^{-S}H{\rm e}^{S}$.  This form may be expanded in
terms of the well-known nested commutator sum.  Another important key
feature of the CCM is that this formally infinite series of nested
commutators actually terminates exactly at terms of second order in
$S$ (i.e., with the double commutator term) for Hamiltonians of the
form of Eq.\ (\ref{H}), as a simple consequence of the basic SU(2)
commutation relations (and see, e.g., Refs.\ \onlinecite{Fa:2004,Ze:1998}
for further details).  A similar exact termination also applies more
generally to the evaluation of the GS expectation value of other
operators of interest, such as the magnetic order parameter, $M$,
discussed below.

The CCM formalism is thus exact if all multispin-slip configurations
are included in the set of set-indices $\{I\}$.  The equations that
need to be solved in practice are coupled sets of nonlinear
(multinomial) equations for the ket-state correlation coefficients
$\{{\cal S}_{I}\}$ and linear equations for the corresponding bra-state
correlation coefficients $\{\tilde{{\cal S}_{I}}\}$, in which the
solutions for $\{{\cal S}_{I}\}$ are needed as input.
Naturally, for practical implementation purposes we will need to make
finite-size truncations of the configurations retained in the GS wave
function, i.e., equivalently, of the set-indices $\{I\}$ retained in
the sums in Eq.\ (\ref{S_operators}).  We will describe below one natural such
systematic truncation hierarchy.  It is important to note that,
since this truncation is the {\it only} approximation made, the CCM in
practice provides a natural series of approximations that provide
systematic improvements in accuracy as one moves to successively higher
levels.

We note that a very important part of the rationale behind the use of
the CCM exponential parametrizations in Eq.\ (\ref{ket_bra_eq}) is
that their use ensures that the method automatically satisfies the
Goldstone linked cluster theorem, even when truncations are made in the
multispin-flip configurations $\{I\}$ retained in the sums in Eq.\ 
(\ref{S_operators}).  Hence, the CCM always obeys size-extensivity at
any level of approximation.  As a consequence the infinite-lattice
(thermodynamic) limit, $N \rightarrow \infty$, may be taken from the
very outset, thereby obviating the need for any finite-size scaling of
the results.  One can also show that at all levels of approximation
the CCM similarly obeys the important Hellmann-Feynman theorem.

Once a suitable approximation hierarchy has been chosen the CCM
equations are derived and solved at successive orders, out to the
highest level that is practically attainable with available
computational resources, as described more fully below.  At each such
order we then calculate the GS energy, $E=\langle\Phi|{\rm
  e}^{-S}H{\rm e}^{S}|\Phi\rangle$, and any other such needed GS
quantity as the average on-site magnetization (or magnetic order
parameter),
$M\equiv-\frac{1}{N}\langle\tilde{\Psi}|\sum^{N}_{i=1}s^{z}_{i}|\Psi\rangle$,
in the rotated spin-coordinates defined on each lattice site, as
described above.  Then, as a final step, we need to extrapolate the
corresponding sequences of approximate results to the exact physical
limit where {\it all} multispin-flip configurations $\{I\}$ are
retained.  We now first describe the approximation scheme used
here, and then describe how the extrapolations are made.

Thus, for our present $s=\frac{1}{2}$ model, we employ the well-known
localized (lattice-animal-based subsystem) LSUB$m$ scheme, which has
by now been very successfully applied to a wide variety of spin-1/2
lattice models.\cite{Fa:2004,Ze:1998,Bishop:1998_J1J2mod,Kr:2000,Bishop:2000,Fa:2001,Darradi:2005,Schm:2006,Bi:2008_PRB_J1xxzJ2xxz,Bi:2008_JPCM,darradi08,Bi:2009_SqTriangle,Darradi:2009_J1J2_XXZmod,richter10:J1J2mod_FM,Bishop:2010_UJack,Bishop:2010_KagomeSq,Reuther:2011_J1J2J3mod,DJJF:2011_honeycomb,Gotze:2011,Bishop:2012_honey_phase,Bishop:2012_checkerboard,Li:2012_honey_full,Bishop:2012_honeyJ1-J2,Li:2012_anisotropic_kagomeSq,RFB:2013_hcomb_SDVBC,Li:2013_chevron}
It is defined such that at the $m$th level of approximation all
possible multispin-flip configurations are retained in the index-set
$\{I\}$ that correspond to locales on the lattice defined by $m$ or
fewer contiguous sites.  Said differently, but equivalently, all
lattice animals of size no larger than $m$ sites are populated with
flipped spins (with respect to the chosen model state $|\Phi\rangle$)
in all possible ways.  Such lattice animals (or contiguous clusters)
are, by definition, contiguous if and only if every site in the
cluster is adjacent (in the NN sense) to at least one other site in
the cluster.  The associated choice of the underlying geometry (or,
perhaps better, topology) of the lattice, i.e., the specification of
which pairs of sites are defined to be NN pairs, also needs to be
made.  There are usually great advantages to making the choice so that
each member of the LSUB$m$ sequence fully respects the underlying
lattice symmetries, as has been explained in more detail elsewhere.\cite{Bishop:2012_checkerboard}  For our present model we hence make, on
physical grounds, the choice that all pairs connected by either
$J_{1}$ bonds or by $J_{2}$ bonds are to be counted as NN pairs.  We
refer henceforth to this definition of NN pairs as the cross-striped
square-lattice geometry.

Even after all space- and point-group symmetries of the lattice and
the particular CCM reference state being used have been incorporated,
the number $N_{f}$ of such {\it distinct} (i.e., under the symmetries)
fundamental configurations retained in an LSUB$m$ approximation
increases very rapidly (usually super-exponentially) with respect to
the truncation index $m$.  Hence, it becomes necessary to use massive
parallelization plus supercomputing
resources\cite{ccm} for
high-order approximations.  In the present study we have been
able to perform LSUB$m$ calculations up to the LSUB10 level for each
of the three classical collinear AFM model states shown in Figs.\
\ref{model_bonds}(a)--(c).  For example, in the cross-striped
square-lattice geometry, $N_{f}=853453$ for the LSUB10 approximation
based on the DN state of Fig.\ \ref{model_bonds}(b) as CCM model
state.  The corresponding numbers at the same LSUB10 level for the
other two model states are slightly smaller but still of the same order of
magnitude.

Finally, we need to extrapolate our LSUB$m$ sequences of
approximations for the GS expectation value of any given operator to
the exact $m \rightarrow \infty$ limit.  For example, although our CCM
LSUB$m$ estimates, $E(m)/N$, do not individually provide upper bounds
for the exact GS energy per spin, $E/N$, due to the corresponding
LSUB$m$ parametrizations of $|\Psi\rangle$ and $\langle\tilde{\Psi}|$
not being manifestly Hermitian conjugates of each other, they do
converge extremely rapidly as $m$ is increased.  We use the very well-tested extrapolation scheme,\cite{Kr:2000,Bishop:2000,Fa:2001,Darradi:2005,Schm:2006,Bi:2008_PRB_J1xxzJ2xxz,darradi08,
  Bi:2008_JPCM,richter10:J1J2mod_FM,Reuther:2011_J1J2J3mod,Bishop:2012_checkerboard,Li:2012_anisotropic_kagomeSq}
\begin{equation}
E(m)/N = a_{0}+a_{1}m^{-2}+a_{2}m^{-4}\,.     \label{E_extrapo}
\end{equation}

Unsurprisingly, the GS expectation values of other physical operators do not converge so rapidly.  For example, the magnetic order parameter $M$ usually obeys a scaling law with leading exponent $1/m$ (rather than $1/m^{2}$ as for the GS energy) for most systems with even moderate amounts of frustration, in which cases an extrapolation scheme of the form
\begin{equation}
M(m) = b_{0}+b_{1}m^{-1}+b_{2}m^{-2}    \label{M_extrapo_standard}
\end{equation}
works well.\cite{Kr:2000,Bishop:2000,Fa:2001,Darradi:2005,DJJF:2011_honeycomb}
On the other hand, for systems either very close to a QCP or for which
the magnetic order parameter of the phase under study is zero or close
to zero, the above extrapolation scheme has been found to overestimate
the magnetic order and to predict a somewhat too large value for the
critical strength of the frustrating interaction that is driving the
transition.  In such cases a scaling law with leading exponent
$1/m^{1/2}$ is found to work much better and we then use the
alternative well-studied extrapolation scheme\cite{Darradi:2005,Schm:2006,Bi:2008_JPCM,Bi:2008_PRB_J1xxzJ2xxz,darradi08,richter10:J1J2mod_FM,Reuther:2011_J1J2J3mod,DJJF:2011_honeycomb,Bishop:2012_checkerboard,Li:2012_anisotropic_kagomeSq}
\begin{equation}
M(m) = c_{0}+c_{1}m^{-1/2}+c_{2}m^{-3/2}\,.   \label{M_extrapo_frustrated}
\end{equation}

Clearly, for the GS expectation value, $Q$, of any physical operator, one may always test for the correct leading exponent $\nu$ in the corresponding LSUB$m$ scaling law,
\begin{equation}
Q(m) = q_{0}+q_{1}m^{-\nu}\,,    
\label{M_extrapo_nu}
\end{equation}
by fitting an LSUB$m$ sequence to this form and treating each of the
parameters $q_{0}$, $q_{1}$, and $\nu$ as fitting parameters.  In
general, of course, any of the above extrapolation schemes of
Eqs.\ (\ref{E_extrapo})--(\ref{M_extrapo_nu}), each with 3 fitting
parameters, is ideally fitted to more than 3 LSUB$m$ data points.

Since the basic square plaquette is such an important structural
element of the lattice, and also since any LSUB$m$ result with $m=2$ is
far from the asymptotic $m \rightarrow \infty$ limit, we prefer to
make any of the LSUB$m$ fits with values $m \geq 4$.  Thus, for most
of the extrapolated results presented in Sec.\ \ref{results_sec} we
use the LSUB$m$ data set $m=\{4,6,8,10\}$.  However, we have also
performed extrapolations using the data sets $m=\{6,8,10\}$, $m=\{4,6,8\}$
and $m=\{2,4,6,8\}$ as a consistency and validity check of our
extrapolations.  For all the GS quantities reported below, we find
extrapolated values which are very insensitive to which data set is used as input.
This both gives credence to our extrapolation schemes and allows us to find a rough
estimate of the inherent error in our quoted results.

For the present model we have performed fits of the form of Eq.\ 
(\ref{M_extrapo_nu}) for the GS energy per spin and for the order
parameter $M$, as reported in Sec.\ \ref{results_sec}.  Similar fits
are reported there too for the susceptibility, $\chi$, which measures
the linear response of the system to various forms of order imposed as
an infinitesimal perturbation to the Hamiltonian.  We discuss in
Sec.\ \ref{results_sec} the corresponding values of the leading
exponent $\nu$ obtained from fits of the form of Eq. 
(\ref{M_extrapo_nu}) for the various calculated GS quantities, and how
they may be used in particular to justify fits for the GS energy and
magnetic order parameter of the form of Eqs.\
(\ref{E_extrapo})--(\ref{M_extrapo_frustrated}) in specific regimes.
We show specifically in some particular cases how the exponent $\nu$ is usually
relatively constant (i.e., only very slowly varying as a function of the
frustration parameter, $\alpha \equiv J_{2}/J_{1}$), except in or very
near critical regimes.

\section{RESULTS AND DISCUSSIONS}
\label{results_sec}
We now present results from our CCM calculations for the spin-1/2
$J_{1}$--$J_{2}$ model on the cross-striped square lattice, whose Hamiltonian
is given by Eq.\ (\ref{H}).  Results are given for the three cases
where the N\'{e}el, double N\'{e}el (DN), and double columnar striped
(DCS) states, shown in Figs.\ \ref{model_bonds}(a), (b), and (c)
respectively, are used in turn as the CCM model states.  In each case
we perform the corresponding LSUB$m$ calculations with $m \leq
10$, as has been discussed in Sec.\ \ref{ccm_sec}.

\begin{figure*}[!tbh]
\begin{center}
\mbox{
\subfigure[]{\scalebox{0.31}{\includegraphics[angle=270]{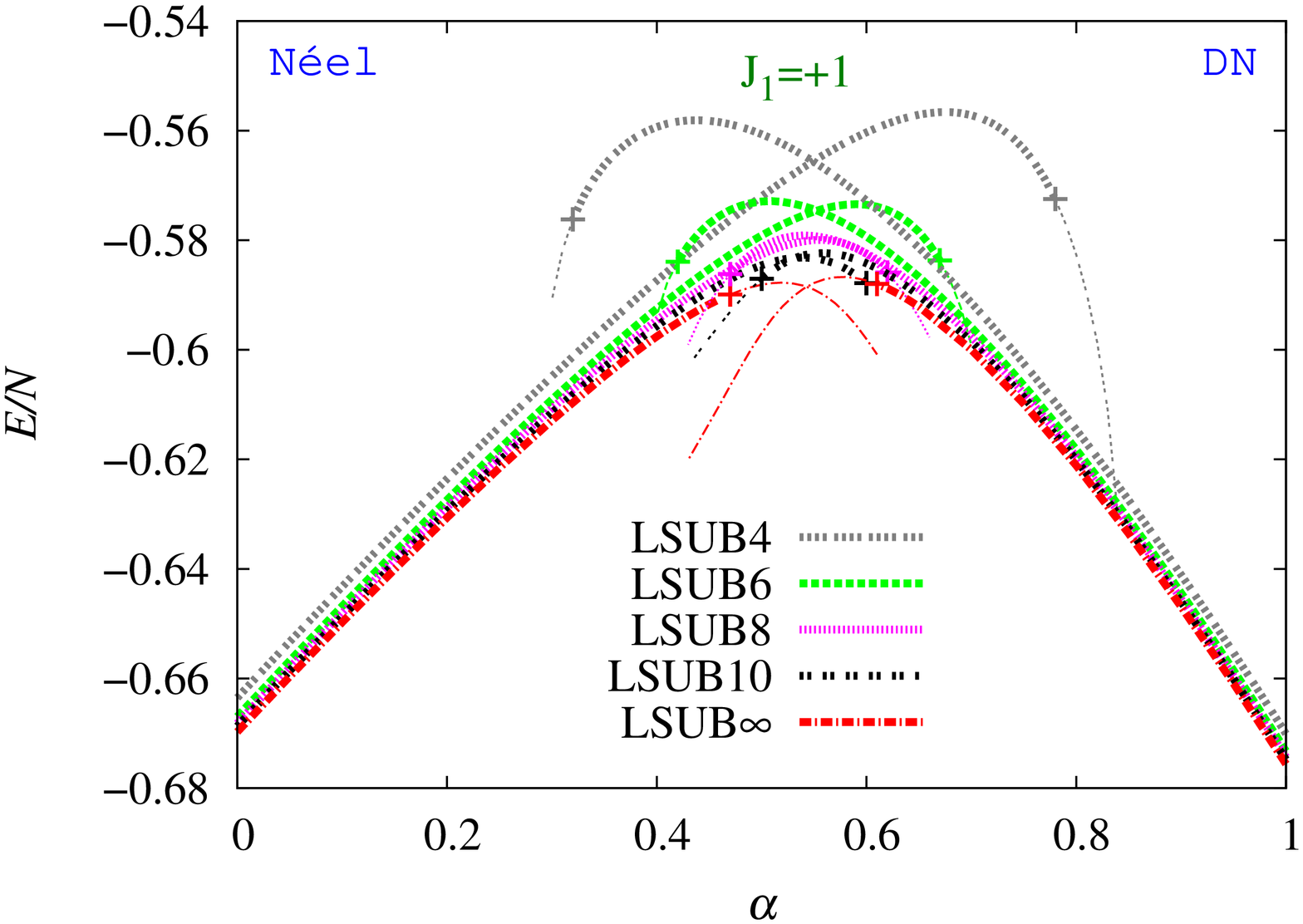}}}
\subfigure[]{\scalebox{0.31}{\includegraphics[angle=270]{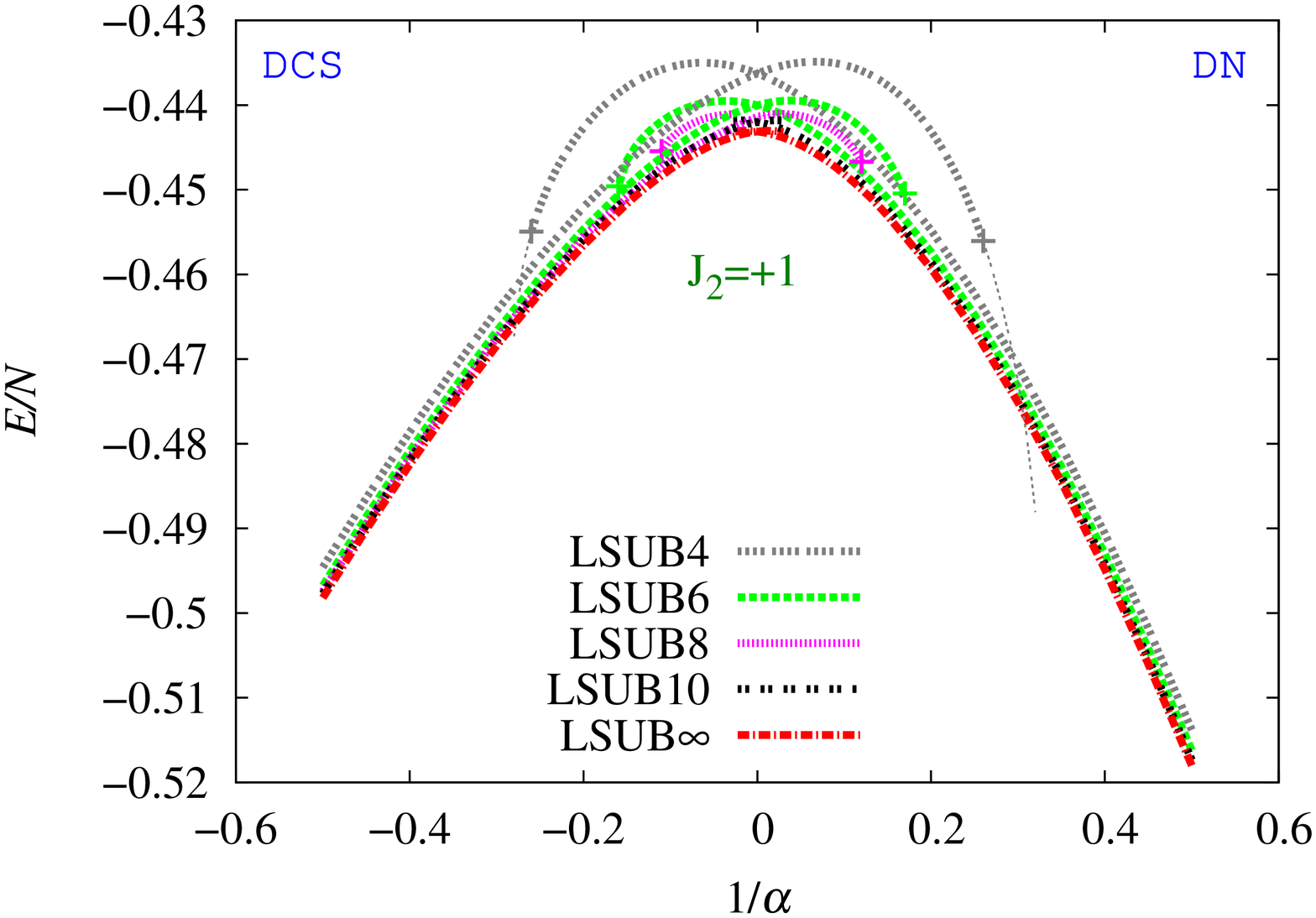}}}
}
\mbox{
\subfigure[]{\scalebox{0.31}{\includegraphics[angle=270]{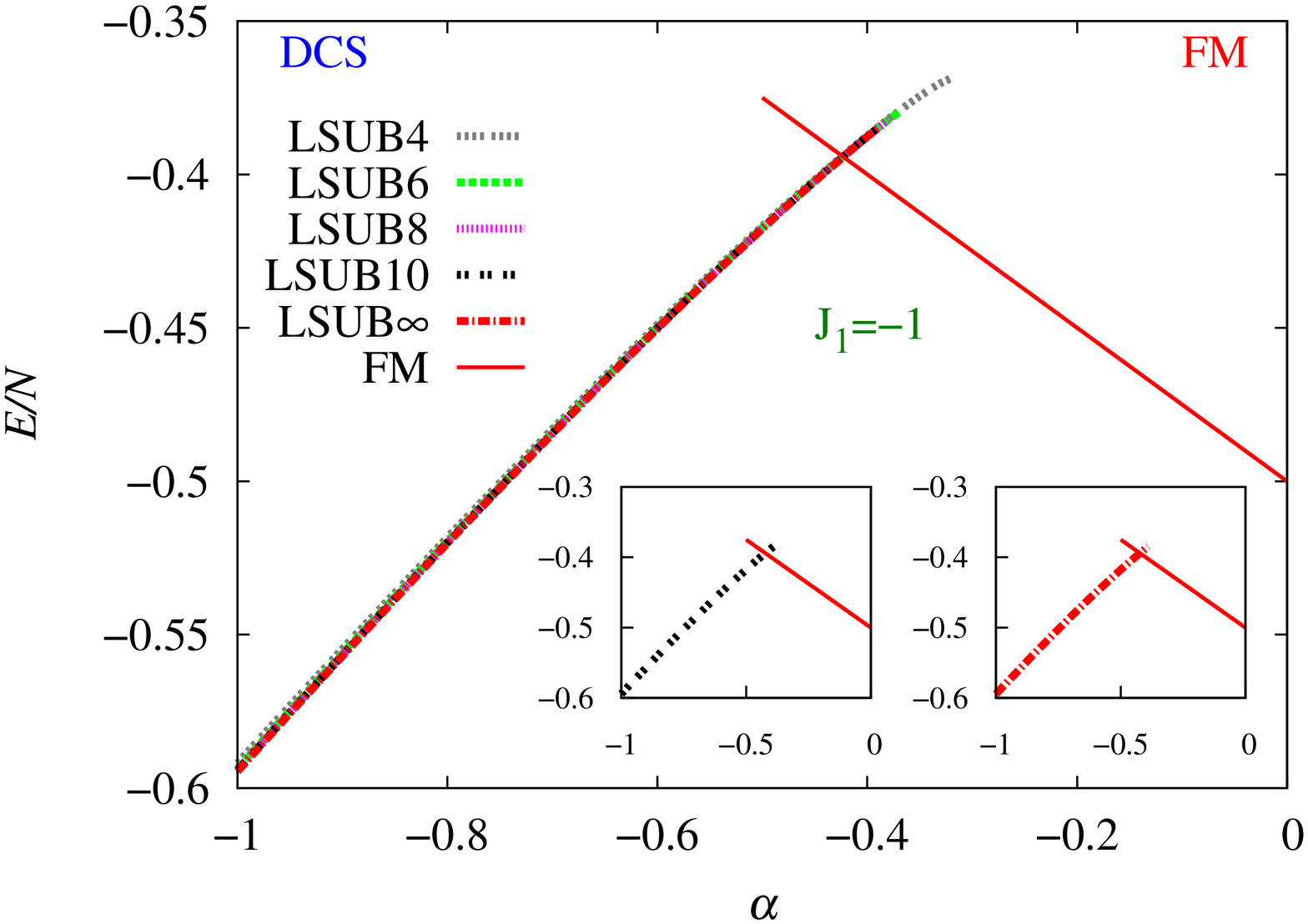}}}
\subfigure[]{\scalebox{0.31}{\includegraphics[angle=270]{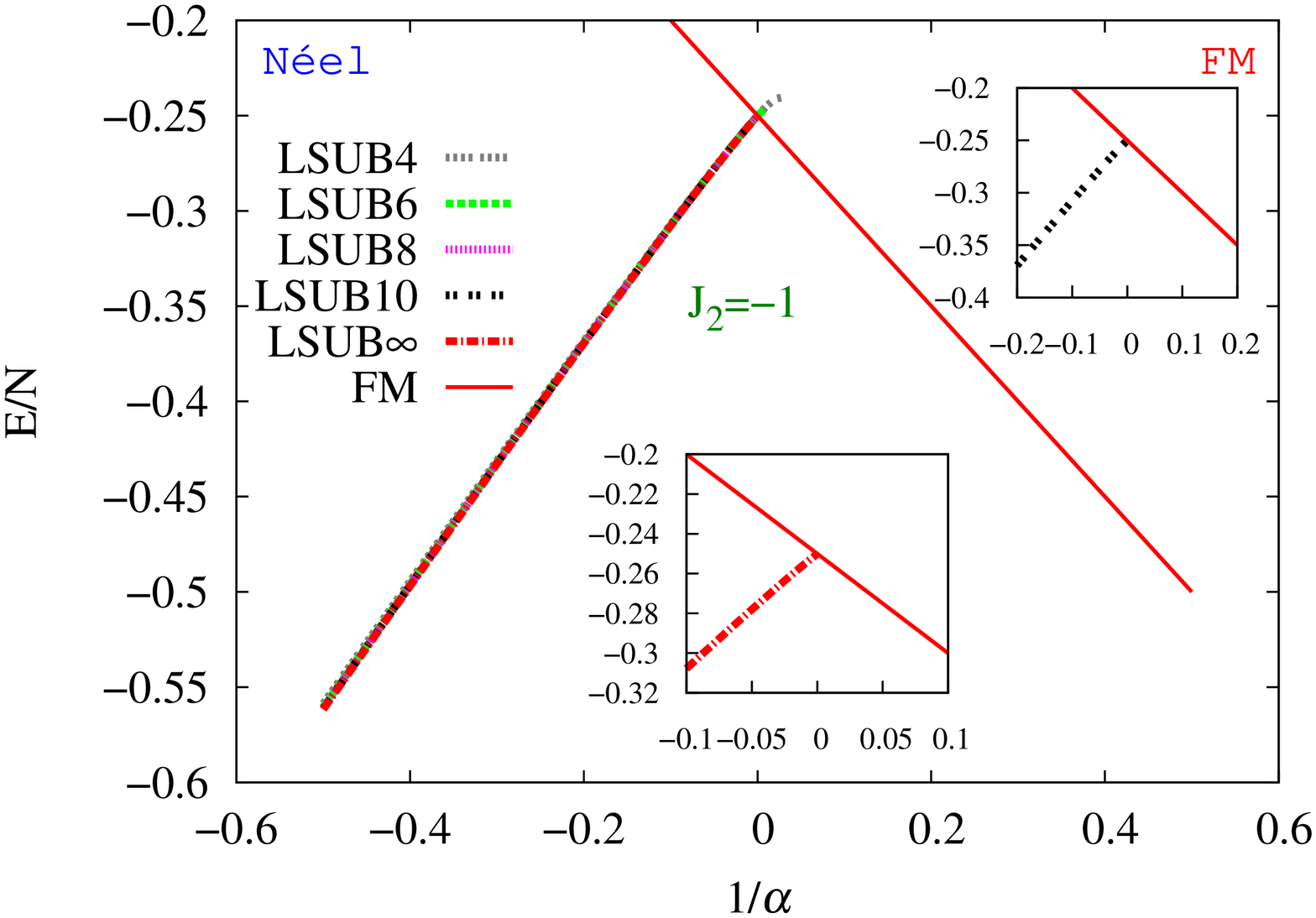}}}
}
\caption{(Color online) CCM LSUB$m$ results with $m=\{4,6,8,10\}$ for the GS energy
  per spin, $E/N$, as a function of the frustration parameter, $\alpha
  \equiv J_{2}/J_{1}$, for the spin-1/2 $J_{1}$--$J_{2}$ model on a
  cross-striped square lattice, for the cases: (a) $J_{1}=+1$ and $0
  \leq \alpha \leq 1$, showing results based on both the N\'{e}el
  (left curves) and DN (right curves) states as model
  states; (b) $J_{2}=+1$ and $-0.5 \leq \alpha^{-1} \leq 0.5$, showing
  results based on both the DCS (left curves) and DN (right curves)
  states as model states; (c) $J_{1}=-1$ and $-1 \leq \alpha
  \leq 0$, showing results based on the DCS state (left curves) as
  model state, as well as the exact FM result (right solid curve),
  $E^{\rm FM}/N = -\frac{1}{4}(2+\alpha)$; and (d) $J_{2}=-1$ and
  $-0.5 \leq \alpha^{-1} \leq 0.5$, showing results based on the
  N\'{e}el state (left curves) as model state, as well as the exact FM
  result (right solid curve), $E^{{\rm FM}}/N=-\frac{1}{4}(1+2/\alpha)$.
  In each case we also show the extrapolated LSUB$\infty$ result
  obtained from Eq.\ (\ref{E_extrapo}) using the data set
  $m=\{4,6,8,10\}$.  All LSUB$m$ solutions are shown out to their
  respective (approximately determined) termination points.  In
  Figs.\ \ref{E}(a) and \ref{E}(b) the plus ($+$) symbols mark the
  points where the respective solutions have $M \rightarrow 0$, and
  those portions of the curves beyond the plus ($+$) symbols shown with
  thinner lines indicate the respective unphysical regions where $M<0$
  (and see text for details).}
\label{E}
\end{center}
\end{figure*}

We first show our CCM results for the GS energy per spin, $E/N$, in Fig.\ \ref{E}, where
we display both LSUB$m$ results with $m=\{4,6,8,10\}$ using each of
the three model states, and the corresponding extrapolated LSUB$\infty$ results
using the scheme of Eq.\ (\ref{E_extrapo}) with this data set.
Firstly, in Fig.\ \ref{E}(a), results for the case $J_{1}>0$ are
presented based on both the N\'{e}el and DN states.  We note that for
both model states results are shown only for certain ranges of the
frustration parameter.  Both sets of curves show a termination point,
an upper one for the N\'{e}el curves and a lower one for the DN
curves.  These CCM LSUB$m$ termination points themselves depend on the
truncation parameter, $m$.  In general we find that the higher is the
index $m$, the smaller is the range of values of $\alpha$ over which
the corresponding (real) GS solution exists based on a particular model
state.

Such terminations of CCM solutions are commonly found and are very
well understood (see, e.g., Refs.\
\onlinecite{Fa:2004,Bi:2009_SqTriangle,Bishop:2010_UJack}).  They are always
reflections of the true quantum phase transitions that are present in
the system under study.  At such termination points the solution to
the corresponding CCM LSUB$m$ equations ceases to be real, and beyond
these points only two unphysical branches of complex conjugate
solutions exist.  On the other hand, in the region before any such
termination point where the true physical solution is real, there
actually must also exist another (unstable) real solution.  Such other
solutions are themselves both unphysical and, fortunately, also very
difficult to determine numerically in general.  In practice any simple
numerical procedure will pick up only the physical branch, which
itself is usually easy to identify by following it (as a function of
the frustration parameter, $\alpha$, for example) to some appropriate
asymptotic limit where it becomes exact or otherwise known.

The two (i.e., the physical and unphysical) real branches of solution
thus meet at a termination point, beyond which they diverge again in
the complex plane as wholly unphysical complex conjugate pairs.  The
values, $\alpha^{t}(m)$ of the termination points for a given branch
of CCM LSUB$m$ solutions, may themselves actually be used to estimate
the corresponding QCP for the GS phase under
study, as $\alpha^{c} = \lim_{m\rightarrow \infty}\alpha^{t}(m)$.
However, it comes as no surprise that the number of iterations
required to solve the CCM LSUB$m$ equations, at a given level of
accuracy, increases significantly as $\alpha \rightarrow
\alpha^{t}(m)$.  Hence, it is computationally expensive to obtain the
values $\alpha^{t}(m)$ with high precision, and since we have accurate other
means available, as described below, to determine the QCPs, we do not
make use of this method here.

Returning to our discussion of Fig.\ \ref{E}(a), we often find (as is
the case here), that for a region near $\alpha^{t}(m)$ on the
corresponding real physical branch the solution itself is also
unphysical in the sense that the corresponding order parameter (here
the local on-site magnetization, $M$) takes negative values.  These
values where $M \rightarrow 0$ (determined as discussed in detail
below) are shown both for the individual LSUB$m$ solutions and the
corresponding LSUB$\infty$ extrapolations as plus ($+$) signs in
Fig.\ \ref{E}(a), and the corresponding regions beyond these points
where $M<0$ are shown with corresponding thinner curves than the
regions marked with thicker curves where $M>0$.  Two points are
particularly noteworthy concerning Fig.\ \ref{E}(a).  Firstly,
whereas the corresponding LSUB$m$ branches of solutions, based on both
the N\'{e}el and DN states as CCM model states, cross at a relatively
sharp angle (as in the classical case, $s \rightarrow \infty$) for
smaller values of the truncation parameter $m$, the angle becomes much
shallower as $m$ increases.  Thus, there are strong preliminary
indications that the counterpart in the $s=\frac{1}{2}$ model of
the classical first-order transition in Fig.\ \ref{model_bonds}(d) at
$\theta^{{\rm cl}}_{1} = \tan^{-1}(\frac{1}{2}$) might become
second-order.  Secondly, it is also apparent from Fig.\ \ref{E}(a)
that the overlap region where CCM LSUB$m$ solutions, for a given value
of $m$, exist for both the N\'{e}el and DN phases becomes smaller as
$m$ increases.  Indeed, for the LSUB$\infty$ extrapolation a clear gap
has opened around $\alpha \approx 0.5$ where neither the N\'{e}el or
the DN phase exists.  We discuss this interesting regime in much
greater detail below.

Before turning to our CCM results based on other model states, it is
worth commenting briefly on the overall accuracy of our results.  To
do so we may, in particular, examine the special case for $J_{1}=+1$
of $\alpha=0$ (i.e., $\theta=0$), corresponding to the N\'{e}el order
of the square-lattice HAF.  Thus, our extrapolated LSUB$\infty$ result
for the GS energy per spin based on our LSUB$m$ results with $m=\{4,6,8,10\}$
and using the N\'{e}el state as CCM model state for this case
$\theta=0$, is $E/N \approx -0.66973$.  This may be compared, for example,
with the corresponding results for the spin-1/2 square-lattice HAF,
$E/N=-0.6693(1)$ from a linked-cluster series expansion technique\cite{Zh:1991}, and $E/N=-0.669437(5)$ from a large-scale QMC simulation,\cite{Sa:1997} free of the usual
``minus-sign-problems'' for this special (unfrustrated limiting) case
where the Marshall-Peierls sign rule\cite{Ma:1955} may be applied.
Our own CCM result is thus in remarkably good agreement with these
benchmark results for this particular case.  We have no reason to
believe that similar accuracy does not pertain over the entire phase
diagram.  Finally, it is worth noting too that our extrapolated result
is extremely robust with respect to the choice of LSUB$m$ data set
used to obtain it.  For example, use of the data sets $m=\{6,8,10\}$
and $m=\{4,6,8\}$ in Eq.\ (\ref{E_extrapo}) yields the
corresponding respective results at $\theta=0$ of $E/N \approx -0.66967$ and
$E/N \approx -0.66977$.  Even inclusion of the very low-order LSUB2 result
with $m=\{2,4,6,8\}$ yields $E/N \approx -0.66981$.

In Fig.\ \ref{E}(b) we show the corresponding energy results for both
the DCS and DN phases in the region $J_{2}=+1$ around
$\theta=\frac{1}{2}\pi$ where they meet in the classical ($s
\rightarrow \infty$) version of the model, as shown in Fig.\
\ref{model_bonds}(d).  Once again, it is clear that the overlap region
where both CCM solutions exist at a given LSUB$m$ level decreases as
the truncation index $m$ increases.  Secondly, just as in Fig.\
\ref{E}(a), the crossing angle of the two curves at $\alpha^{-1}=0$
becomes much shallower as $m$ increases, again more indicative of a
continuous (second-order) transition than the corresponding
first-order transition in the classical ($s \rightarrow \infty$)
version of the model.

The crossing point at $J_{1}=0$ (with $J_{2}=+1$) of each of the pairs
of LSUB$m$ curves based on the DCS and DN states as CCM model states
is precisely the limiting case of decoupled 1D HAF $J_{2}$ chains.
Hence, it is again interesting to ascertain the accuracy of our
results by comparison with the exact results in this soluble limit.
Thus, our extrapolated LSUB$\infty$ result for the GS energy per spin,
based on either the DCS or DN model state, for this case $\theta =
\frac{1}{2}\pi$ (with $J_{2}=+1$), and using the extrapolation scheme
of Eq.\ (\ref{E_extrapo}) with $m=\{4,6,8,10\}$, is
$E/N \approx -0.44312$.  Again, our results are remarkably robust with respect
to the choice of LSUB$m$ data set used.  Thus, for example, use of the
data sets $m=\{6,8,10\}$ and $m=\{4,6,8\}$ in Eq.\ (\ref{E_extrapo}) yields the corresponding results at
$\theta=\frac{1}{2}\pi$ of $E/N \approx -0.44313$ and $E/N \approx -0.44311$,
respectively.  Even inclusion of the very low-order LSUB2 result with
$m=\{2,4,6,8\}$ yields $E/N \approx -0.44308$.  Thus, once again, our CCM
results are seen to be in excellent agreement with the corresponding
exact result, $E/N=\frac{1}{4}-\ln 2 \approx -0.44315$, from the Bethe
ansatz solution.\cite{Bethe:1931,Hulthen:1938}

Let us now turn to the case $J_{1}=-1$.  In Fig.\ \ref{E}(c) we show
our CCM results based on the DCS state as model state in this region.
We note first that over the entire regime shown the LSUB$m$ results
converge extremely rapidly as the order $m$ increases.  Secondly, we
note too that the LSUB$m$ termination points also similarly converge
rather fast, and approach the crossing point with the exact FM
eigenstate.  This is explicitly shown in the inset for the LSUB10
case.  The crossing point of the LSUB$\infty$ DCS curve with the FM
curve is now at the value $\alpha = -0.423(1)$, irrespective of
which LSUB$m$ data set is used to perform the extrapolation.

With respect to the GS energy, finally we show in Fig.\ \ref{E}(d)
our results based on the N\'{e}el state in the unfrustrated region
where $J_{2}<0$.  As in the previous case of Fig.\ \ref{E}(c), the
LSUB$m$ results again converge very rapidly as the truncation order
parameter $m$ increases.  Similarly too, the LSUB$m$ termination
points converge very rapidly to precisely the point $J_{1}=0$ where
all of our energy results cross that of the exact FM eigenstate, which
is also shown in Fig.\ \ref{E}(d), as can be explicitly seen in the
inset to the figure for the LSUB10 case.

\begin{figure*}[!tbh]
\begin{center}
\mbox{
\subfigure[]{\scalebox{0.31}{\includegraphics[angle=270]{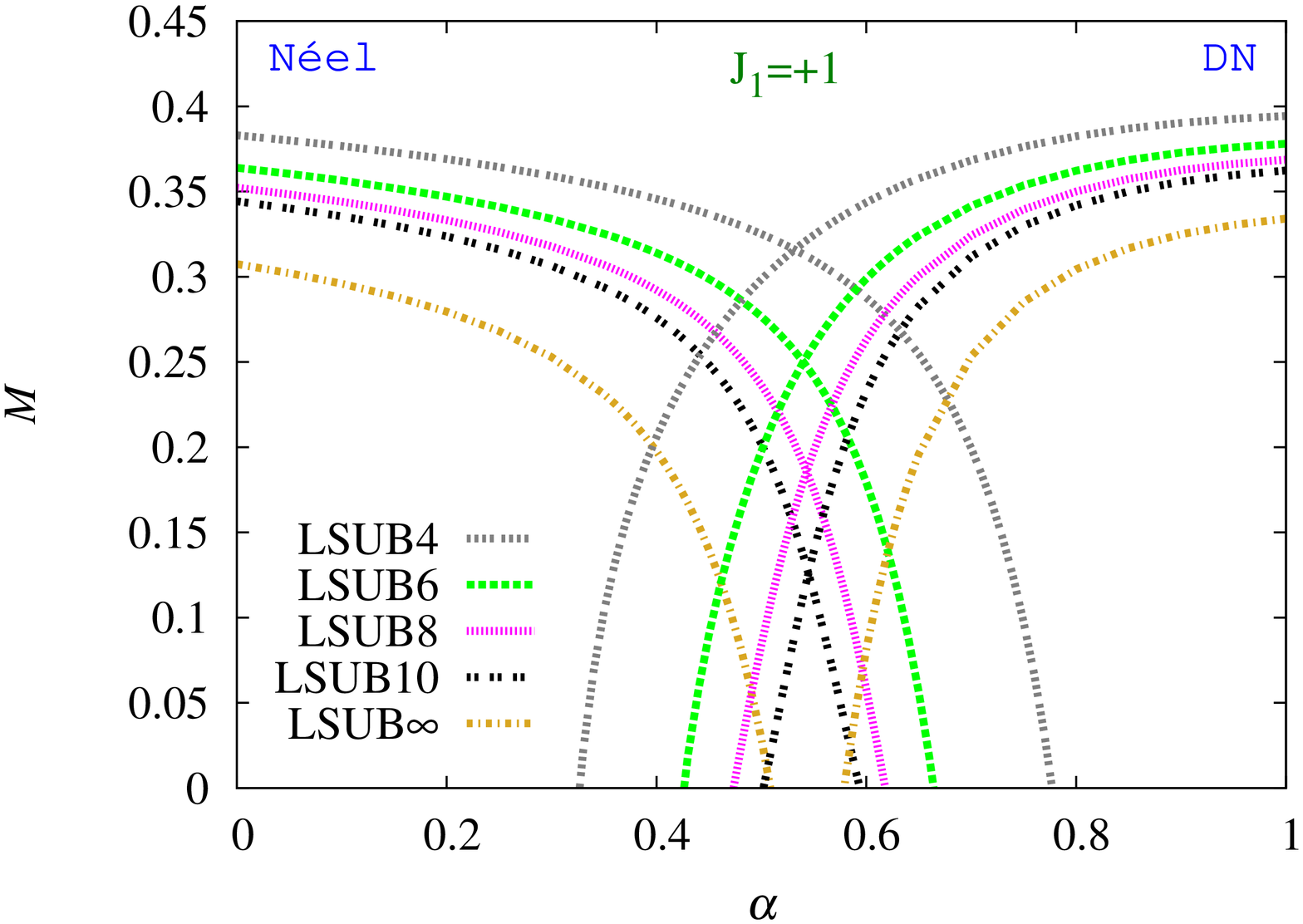}}}
\subfigure[]{\scalebox{0.31}{\includegraphics[angle=270]{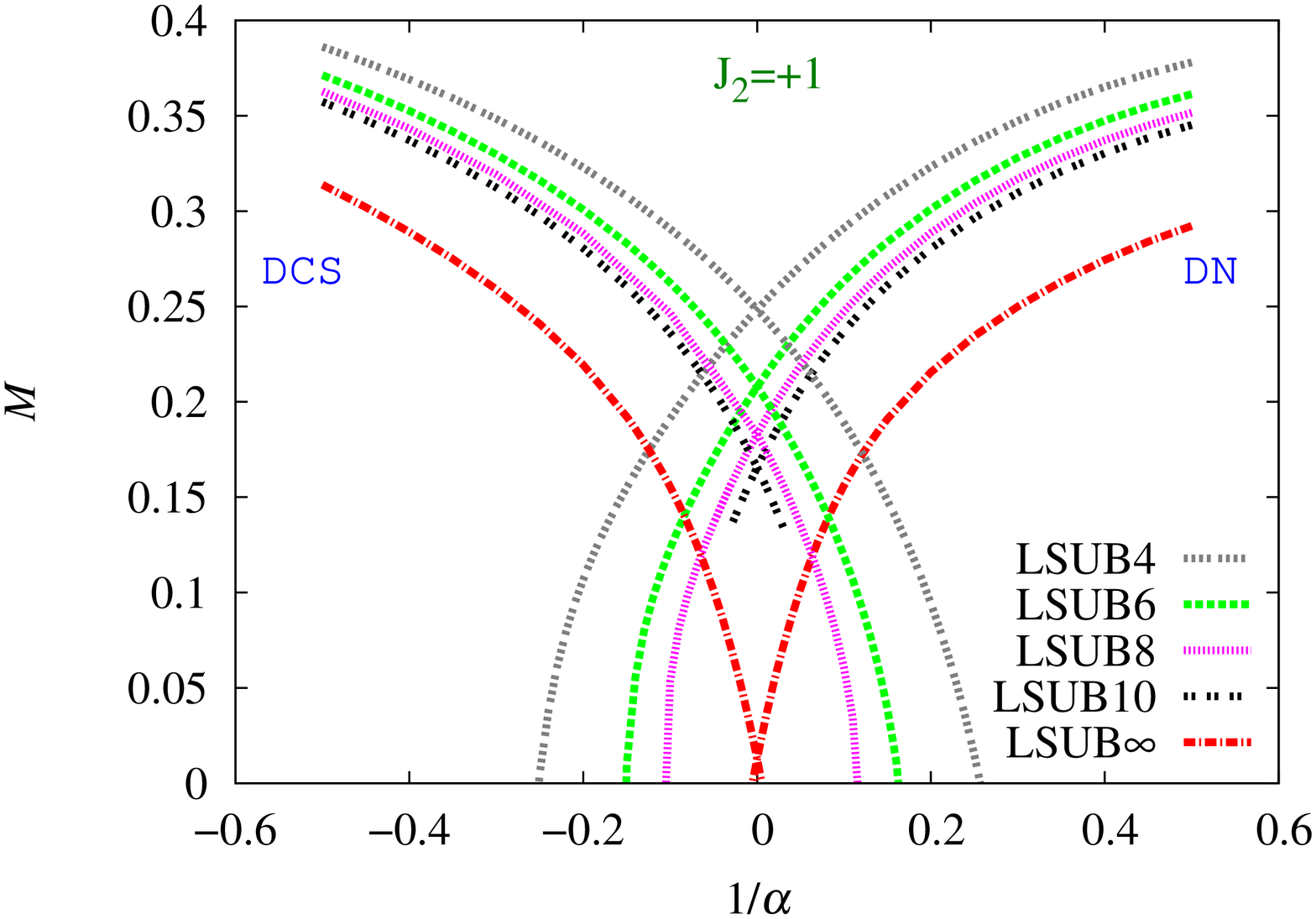}}}
}
\mbox{
\subfigure[]{\scalebox{0.31}{\includegraphics[angle=270]{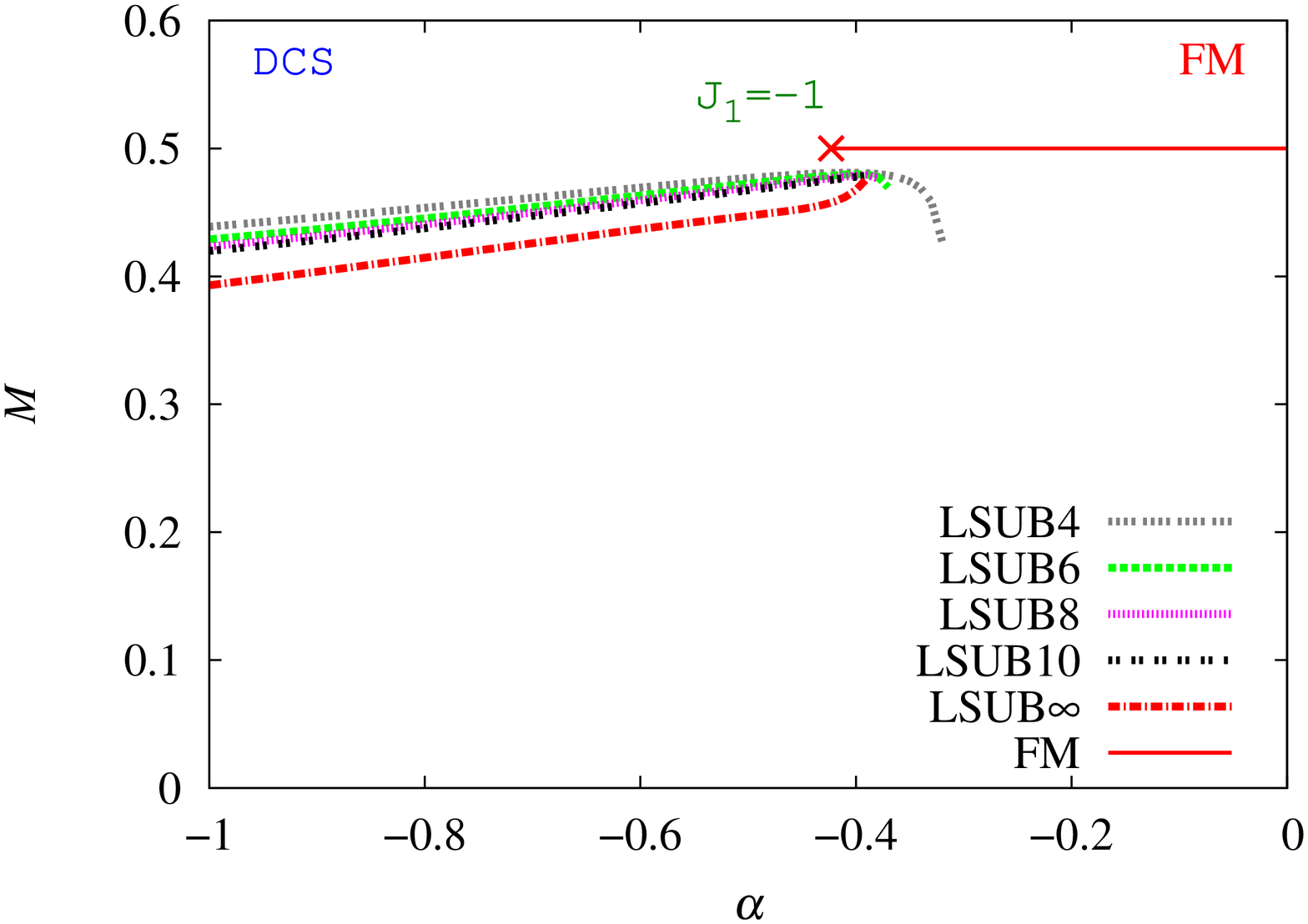}}}
\subfigure[]{\scalebox{0.31}{\includegraphics[angle=270]{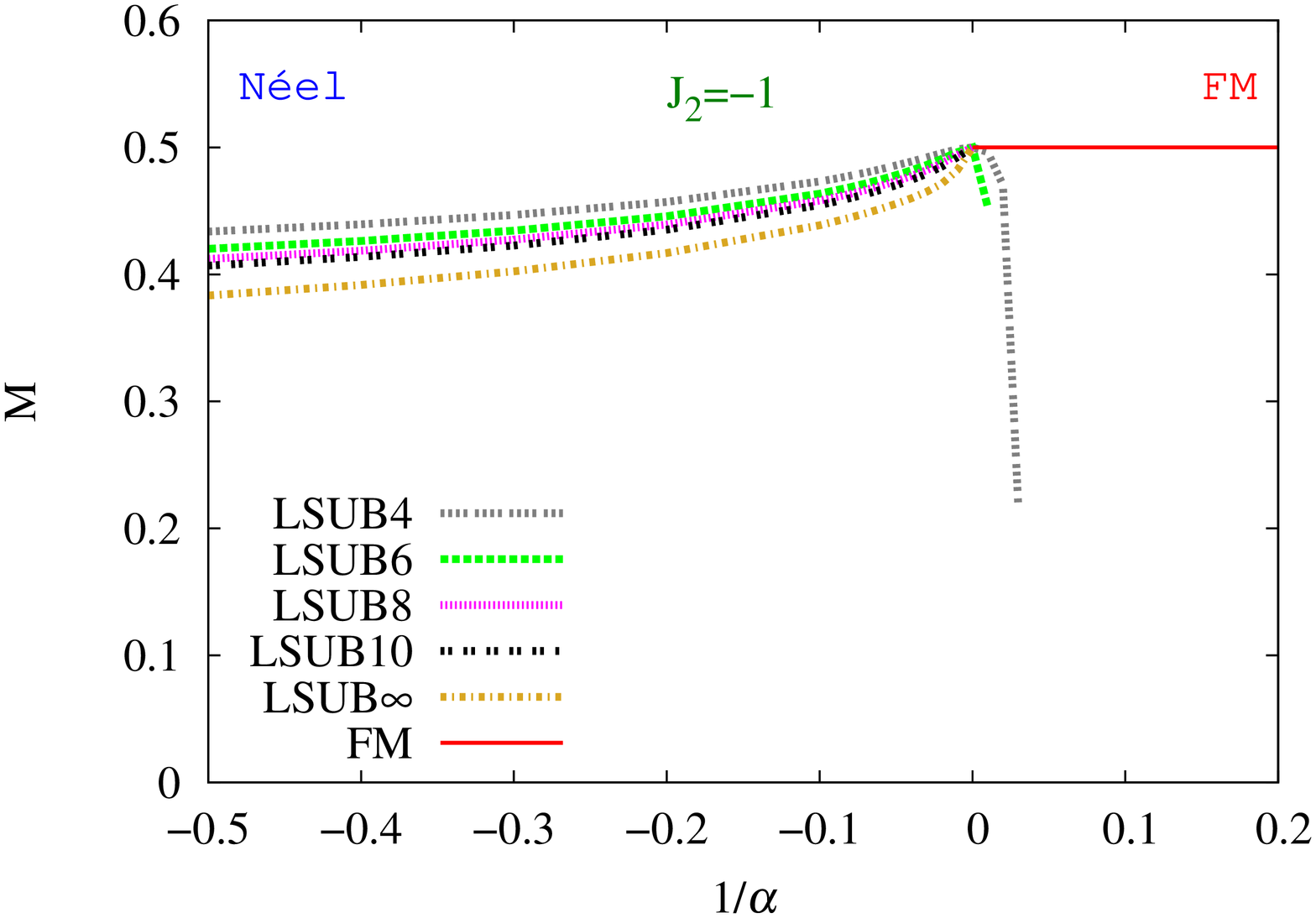}}}
}
\caption{(Color online) CCM LSUB$m$ results with $m=\{4,6,8,10\}$ for the magnetic
  order parameter, $M$, as a function of the frustration parameter,
  $\alpha \equiv J_{2}/J_{1}$, for the spin-1/2 $J_{1}$--$J_{2}$ model
  on a cross-striped square lattice, for the cases: (a) $J_{1}=+1$ and
  $0 \leq \alpha \leq 1$, showing results based on both the N\'{e}el
  (left curves) and DN (right curves) states as model states; (b)
  $J_{2}=+1$ and $-0.5 \leq \alpha^{-1} \leq 0.5$, showing results
  based on both the DCS (left curves) and DN (right curves) states as
  model states; (c) $J_{1}=-1$ and $-1 \leq \alpha \leq 0$, showing
  results based on the DCS state (left curves) as model state, as well
  as the exact FM result (right solid curve), $M^{\rm FM} =
  \frac{1}{2}$; and (d) $J_{1}=-1$ and $-0.5 \leq \alpha^{-1} \leq
  0.2$, showing results based on the N\'{e}el state (left curves) as
  model state, as well as the exact FM result (right solid curve),
  $M^{{\rm FM}}=\frac{1}{2}$.  In each case we also show the
  extrapolated LSUB$\infty$ result obtained by inputting the data set
  $m=\{4,6,8,10\}$ into Eq.\ (\ref{M_extrapo_standard}) for cases
  (a) and (d) and into Eq.\ (\ref{M_extrapo_frustrated}) for cases
  (b) and (c).  The cross ($\times$) symbol in Fig.\ \ref{M}(c) marks the
  position of the corresponding energy crossing point in Fig.\
  \ref{E}(c).}
\label{M}
\end{center}
\end{figure*}

To summarize our results obtained from the energy calculations, we
have found strong definite evidence so far of five QCPs, four in the
frustrated region where $J_{2}>0$ and one in the unfrustrated region
where $J_{2}<0$.  Firstly, in the (frustrated) first quadrant of the
phase diagram where $J_{1}>0$ and $J_{2}>0$, the classical critical
point at $\alpha^{{\rm cl}}_{1} = \frac{1}{2}$ appears to be split
into two QCPs in the $s=\frac{1}{2}$ case at positions
$\alpha^{c}_{1a} \lesssim 0.5$ and $\alpha^{c}_{1b} \approx 0.6$, with
a N\'{e}el-ordered GS phase for $\alpha < \alpha^{c}_{1a}$, a
DN-ordered GS phase for $\alpha > \alpha^{c}_{1b}$, and an as yet
unknown intermediate phase.  Secondly, we find that the spin-1/2 and
classical versions of the model share a common critical point at
$J_{1}=0$ when $\theta^{c}_{2}=\frac{1}{2}\pi=\theta^{{\rm cl}}_{2}$,
at which the DN-ordered GS phase for values $J_{1}>0$ yields to the
DCS-ordered GS phase for values $J_{1}<0$.  However, unlike the
classical first-order transition at this point, its $s=\frac{1}{2}$
quantum analog seems to be more second-order in character in terms of the
energy results.  Thirdly, in the (frustrated) second quadrant of the
phase diagram where $J_{1}<0$ and $J_{2}>0$, the classical critical
point at $\alpha^{{\rm cl}}_{3}=-\frac{1}{2}$, at which the
DCS-ordered GS phase yields to the FM-ordered GS phase, is shifted in
the $s=\frac{1}{2}$ case to a QCP at $\alpha^{c}_{3}=-0.423(1)$.
Finally, in the (unfrustrated) lower hemisphere of the phase diagram
where $J_{2}<0$, we find, as expected, that the spin-1/2 and classical versions of
the model share a common critical point at $J_{1}=0$ when
$\theta^{c}_{4}=\frac{3}{2}\pi=\theta^{{\rm cl}}_{4}$ at which the
FM-ordered GS phase for values $J_{1}<0$ yields to the
N\'{e}el-ordered GS phase for values $J_{1}>0$.

In order to examine the nature of these QCPs in more detail, and
especially the positions of the two QCPs at $\alpha^{c}_{1a}$ and
$\alpha^{c}_{1b}$, we now turn our attention to our corresponding
results for the GS order parameter $M$.  Our LSUB$m$ results with
$m=\{4,6,8,10\}$ using each of the previous CCM model states are shown
in Fig.\ \ref{M}.

Figure \ref{M}(a) presents the analogous results
for $M$ based on both the N\'{e}el and DN states as shown in Fig.\
\ref{E}(a) for the GS energy, applicable to the (frustrated) first
quadrant of the phase diagram with $J_{1}>0$ and $J_{2}>0$.  The plus
($+$) symbols shown in Fig.\ \ref{E}(a) for the LSUB$m$ results
presented there correspond to the respective points in Fig.\
\ref{M}(a) at which $M=0$.

\begin{figure*}[!t]
\begin{center}
\mbox{
\subfigure[]{\scalebox{0.31}{\includegraphics[angle=270]{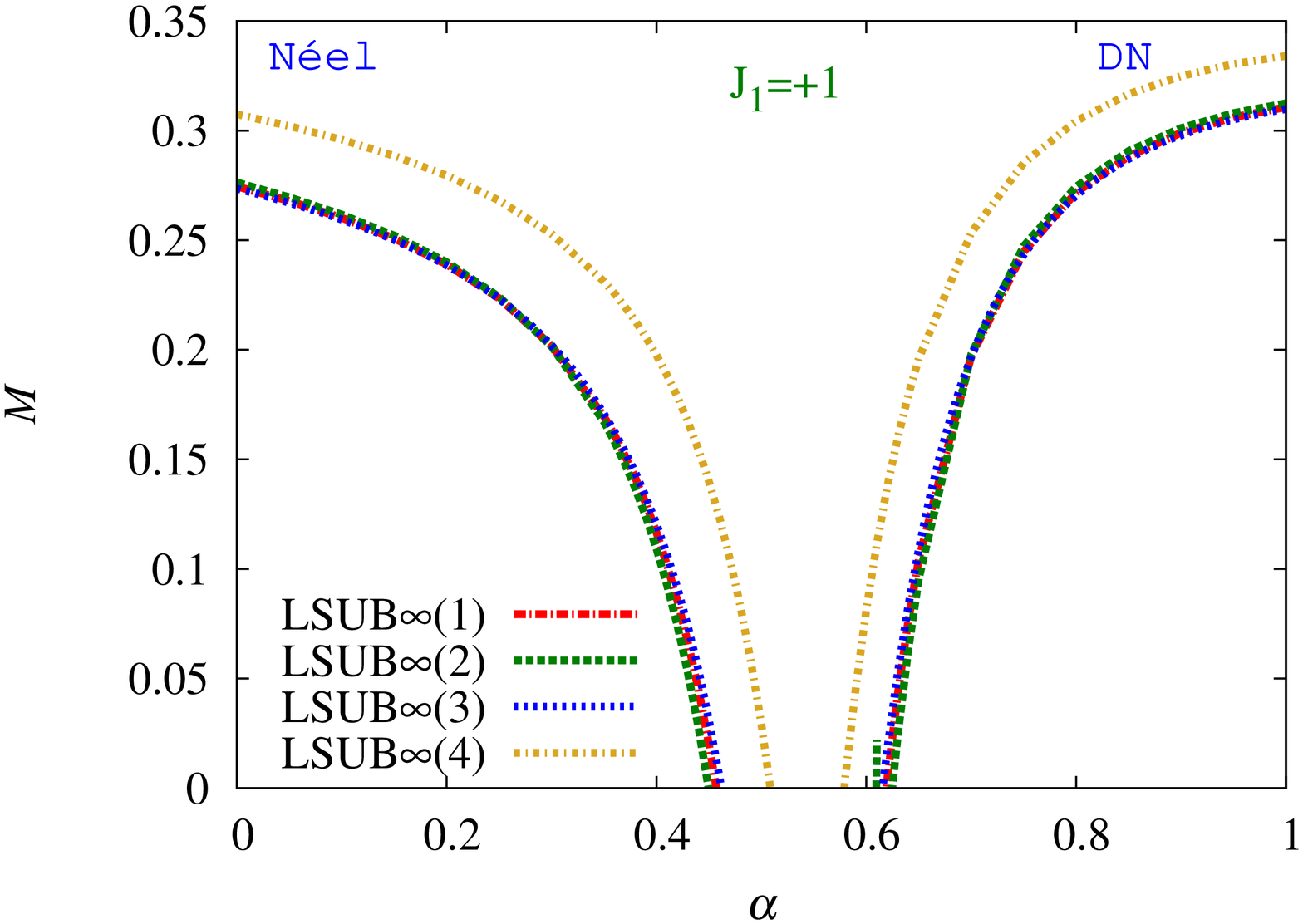}}}
\subfigure[]{\scalebox{0.31}{\includegraphics[angle=270]{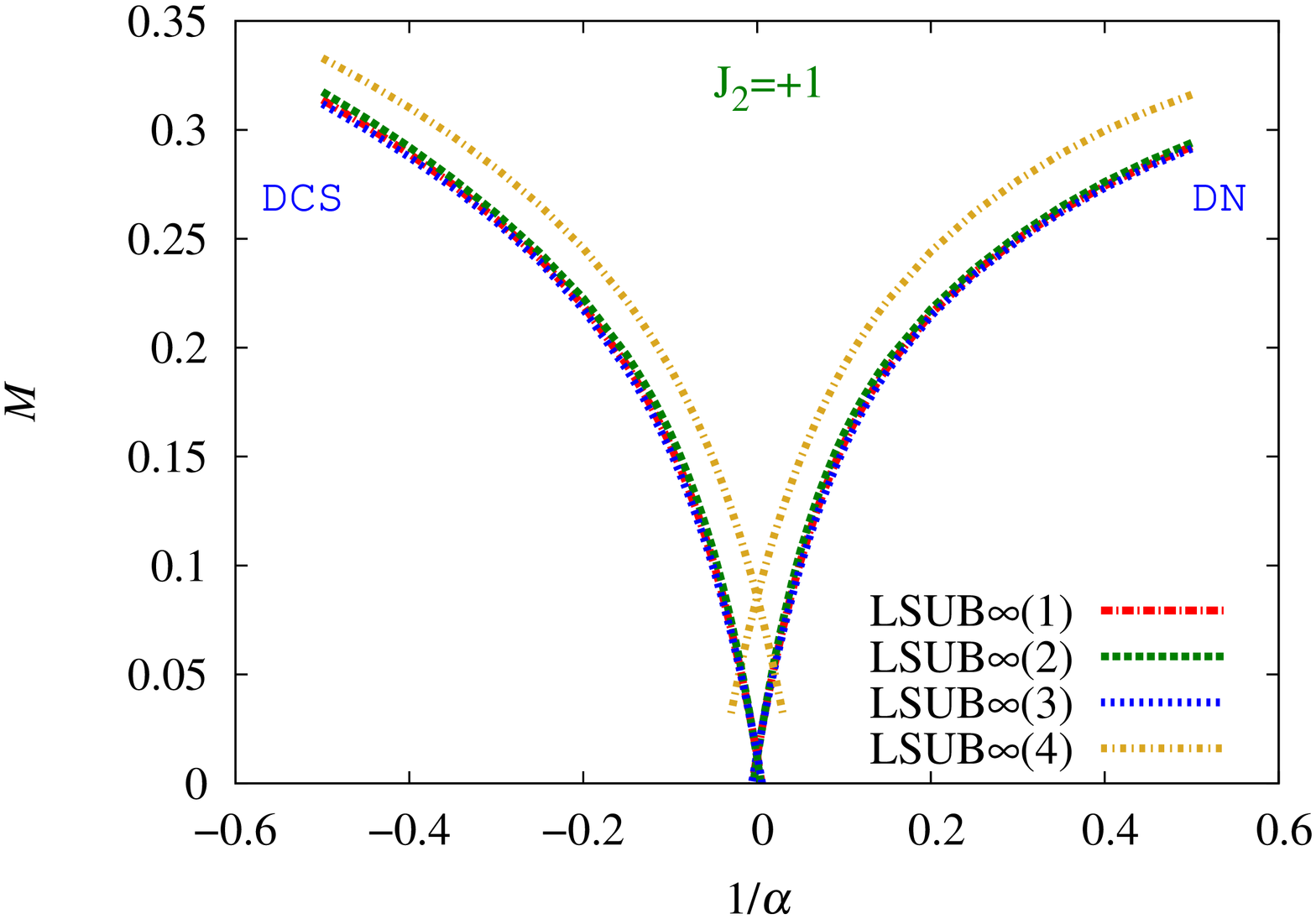}}}
}
\caption{(Color online) Various extrapolated CCM LSUB$\infty$ results for the magnetic order
  parameter, $M$, as a function of the frustration parameter, $\alpha \equiv J_2/J_1$, for the
  spin-1/2 $J_{1}$--$J_{2}$ model on a cross-striped square lattice, for the cases: (a) $J_{1}=+1$ and $0
  \leq \alpha \leq 1$, showing results based on both the N\'{e}el
  (left curves) and DN (right curves) states as model
  states; (b) $J_{2}=+1$ and $-0.5 \leq \alpha^{-1} \leq 0.5$, showing
  results based on both the DCS (left curves) and DN (right curves)
  states as model states.  The curves LSUB$\infty(k)$ with $k=1,2,3$ use the
  extrapolation scheme of Eq.\ (\ref{M_extrapo_frustrated}) with LSUB$m$
  data sets respectively: $k=1$, $m=\{4,6,8,10\}$; $k=2$,
  $m=\{6,8,10\}$; and $k=3$, $m=\{4,6,8\}$; while the curve LSUB$\infty(4)$ uses the extrapolation scheme
  of Eq.\ (\ref{M_extrapo_standard}), with LSUB$m$ data set
  $m=\{4,6,8,10\}$.}
\label{M_Extrapolation}
\end{center}
\end{figure*}

In order to consider again the special case for $J_{1}=+1$ of
$\alpha=0$, corresponding to the N\'{e}el order of the square-lattice
HAF, we also show in Fig.\ \ref{M}(a) the extrapolated LSUB$\infty$
result for the order parameter $M$ based on our LSUB$m$ results for
the N\'{e}el model state with $m=\{4,6,8,10\}$ used in the scheme of
Eq.\ (\ref{M_extrapo_standard}), which is applicable for this
unfrustrated limiting case.  Our corresponding estimate for the square-lattice HAF is then $M\approx 0.307$.  This may again be compared with the
corresponding result $M=0.307(1)$ from a linked-cluster series
expansion technique,\cite{Zh:1991} and $M=0.3070(3)$ from a
large-scale quantum Monte Carlo simulation.\cite{Sa:1997}  Once again
we may demonstrate the robustness of our extrapolation by comparing
results obtained from the use of different data sets.  For
example, use of the data sets $m=\{6,8,10\}$ and $m=\{4,6,8\}$ in
Eq.\ (\ref{M_extrapo_standard}) yields the corresponding respective
results at $\theta=0$ of $M\approx 0.305$ and $M \approx 0.309$.  Even inclusion of the very low-order LSUB2 result into the set $m=\{2,4,6,8\}$ still
  gives the extremely good result $M\approx 0.314$.

  We have shown the LSUB$\infty$ extrapolated values for $M$ in Fig.\
  \ref{M}(a) using the scheme of Eq.\ (\ref{M_extrapo_standard}),
  since we wished primarily to use it to determine the accuracy of
  our technique at the special unfrustrated point $\alpha=0$ where
  this scheme is especially appropriate.  However, when we now turn
  our attention to the very interesting QCPs at $\alpha^{c}_{1a}$ and
  $\alpha^{c}_{1b}$ the extrapolation scheme of Eq.\ 
  (\ref{M_extrapo_standard}) loses its validity, and instead the
  scheme of Eq.\ (\ref{M_extrapo_frustrated}) becomes apposite.
  Nevertheless, Fig.\ \ref{E}(a) shows clearly that even use of the
  scheme of Eq.\ (\ref{M_extrapo_standard}) gives clear
  indications of a gap between the N\'{e}el and DN phases, which can
  only widen when the more appropriate scheme of Eq.\ 
  (\ref{M_extrapo_frustrated}) is used in this critical regime.

  Thus, in Fig.\ \ref{M_Extrapolation}(a) we now show the corresponding
  extrapolated results using the scheme of Eq.\ (\ref{M_extrapo_frustrated}), and
  where we also demonstrate the robustness of our fitting procedure by
  using various LSUB$m$ data sets to perform the fits.  
The use of
  such a sensitivity analysis as shown in Fig.\
  \ref{M_Extrapolation}(a) yields values for the corresponding QCPs,
  $\alpha^{c}_{1a} = 0.46(1)$ and $\alpha^{c}_{1b} = 0.62(1)$.

  The results shown in both Figs.\ \ref{M}(b) and
  \ref{M_Extrapolation}(b) also show very clearly the phase transition
  between the DCS and DN phases at $\theta^{c}_{2}=\frac{1}{2}\pi$.  In
  particular, Fig.\ \ref{M_Extrapolation}(b) demonstrates that when
  the extrapolation scheme of Eq.\ (\ref{M_extrapo_frustrated}) is
  used, as is appropriate at the QCP, the order parameter $M$ becomes
  zero within extremely small error bars on both sides of the
  transition precisely at the QCP, thereby adding considerable weight
  to the conclusion from the GS energy results that this transition is
  a (continuous) second-order one, quite unlike its classical
  first-order counterpart.

  It is worth emphasizing that, although we show in
  Fig.~\ref{M_Extrapolation}(b) extrapolations based on both Eqs.\
  (\ref{M_extrapo_standard}) and (\ref{M_extrapo_frustrated}), for the
  sake of comparison and completeness, the proper choice in this case
  is most definitely Eq.~(\ref{M_extrapo_frustrated}) for reasons
  stated above and in Sec.~\ref{ccm_sec}.  Furthermore, as we have
  indicated, in any such analysis we may also use Eq.\ (\ref{M_extrapo_nu})
  for a first fit to the results, in order to find the leading
  exponent.  In the case of the results shown in Fig.\
  \ref{M_Extrapolation}(b), for example, such a fit clearly shows that
  Eq.\ (\ref{M_extrapo_frustrated}) is indeed the appropriate choice,
  fully as expected from much accumulated prior experience.

  In Fig.\ \ref{M}(c) we show the corresponding CCM results for the
  order parameter to those shown in Fig.\ \ref{E}(c) for the GS
  energy, in the region where the DCS and FM phases meet.  As
  discussed previously, the LSUB$m$ results based on the DCS state as
  model state terminate at points, depending on the truncation
  parameter $m$, that always extend slightly into the region where the
  FM state is the stable phase, but where the nonphysical region
  decreases as $m$ increases.  The DCS termination point for the
  LSUB10 approximation is, for example, at a value $\alpha \approx
  -0.39$, and the corresponding extrapolated LSUB$\infty$ value, shown
  in Fig.\ \ref{M}(c), based on the extrapolation scheme of Eq.\
  (\ref{M_extrapo_frustrated}), also terminates at this value.  For
  comparison, the cross ($\times$) symbol in Fig.\ \ref{M}(c) on the
  FM curve, $M^{{\rm FM}}=\frac{1}{2}$, marks the position, $\alpha
  \approx -0.423$, of the corresponding energy crossing point from
  Fig.\ \ref{E}(c).  It seems clear that if we could go to arbitrarily
  high LSUB$m$ orders in this case the DCS order parameter would
  approach the value 0.5 with a similar cusp shape as in our
  approximate LSUB$\infty$ result in Fig.\ \ref{M}(c) at precisely the
  energy crossing point, namely $\alpha^{c}_{3}$.  For this particular
  transition, the energy results clearly give a more accurate estimate
  for $\alpha^{c}_{3}$ than the order parameter results.

  Finally, with respect to the magnetic order parameter, we show in
  Fig.\ \ref{M}(d) the corresponding CCM results for the N\'{e}el
  phase in the region where it meets the FM phase.  The appropriately
  extrapolated LSUB$\infty$ result shows clearly how $M$ approaches
  the value 0.5 on the N\'{e}el side with a similar cusp to that
  observed in Fig.\ \ref{M}(c).

  Clearly, our results for $M$ completely reinforce the conclusions we
  have already drawn from our corresponding results for the GS energy.
  Taken together they give clear evidence for the quantum
  $s=\frac{1}{2}$ model to contain five phases in the GS phase
  diagram, by contrast with the four phases of its classical ($s
  \rightarrow \infty$) counterpart.  We have also found accurate
  values for all five QCPs.  What remains unclear up till now,
  however, is the nature of the phase in the regime $\alpha^{c}_{1a} <
  \alpha < \alpha^{c}_{1b}$.  In order to shed light on this remaining
  issue we now investigate the susceptibility of our CCM solutions in
  this regime to various forms of valence-bond crystalline (VBC)
  order.

\begin{figure}
\begin{center}
\subfigure[]{\scalebox{0.4}{\includegraphics{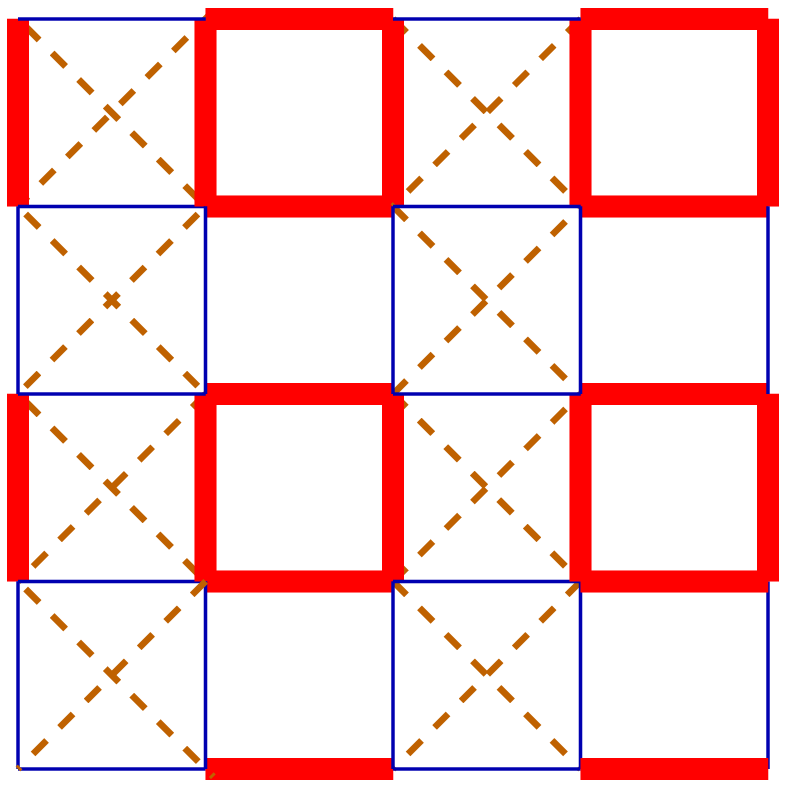}}} \qquad
\subfigure[]{\scalebox{0.4}{\includegraphics{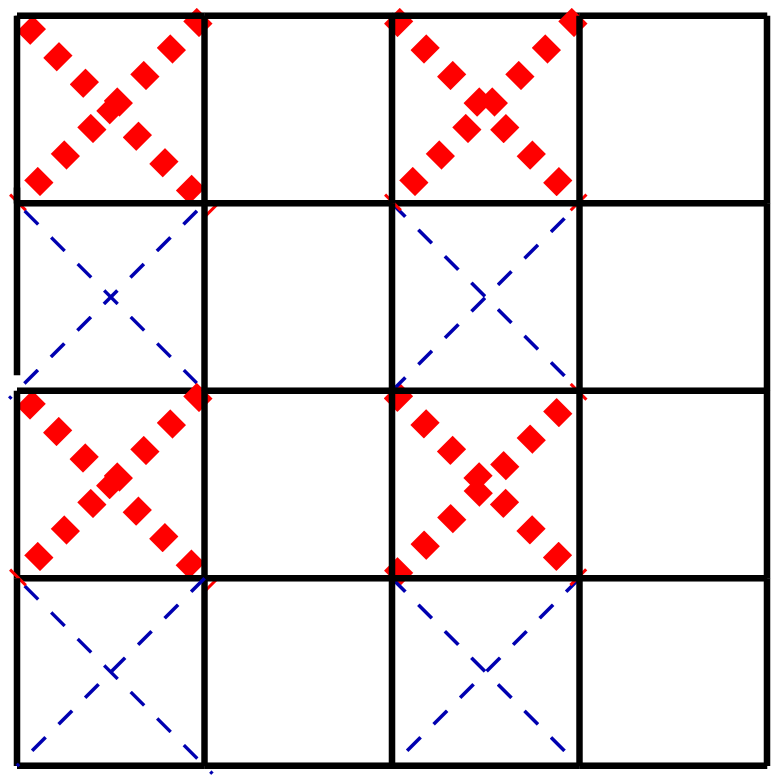}}} 
\caption{(Color online) The fields (or perturbations), $F=\delta\; \hat{O}$, for the
  two forms of valence-bond crystalline (VBC) susceptibility
  considered, namely: (a) plaquette (PVBC), $\chi_{p}$; and (b)
  crossed-dimer (CDVBC), $\chi_{d}$.  In case (a) the thick (red) and
  thin (blue) solid lines correspond respectively to strengthened and
  weakened $J_{1}$ exchange couplings, where $\hat{O}_{p} =
  \sum_{\langle i,j \rangle} a_{ij}
  \mathbf{s}_{i}\cdot\mathbf{s}_{j}$, and the sum runs over all NN
  bonds of the square lattice, with $a_{ij}=+1$ and $-1$ for thick
  (red) and thin (blue) lines respectively, as shown.  In case (b) the
  thick (red) and thin (blue) dashed lines correspond respectively to
  strengthened and weakened $J_{2}$ exchange couplings, where
  $\hat{O}_{d} = \sum_{\langle\langle i,k \rangle\rangle'} a_{ik}
  \mathbf{s}_{i}\cdot\mathbf{s}_{k}$, and the sum runs over all
  diagonal bonds of the cross-striped square lattice, with $a_{ik}=+1$
  and $-1$ for thick (red) and thin (blue) dashed lines respectively.  The
  original solid (black) $J_{1}$ bonds and dashed (brown) $J_{2}$
  bonds are unaltered in strength.}
\label{VBC_patterns}
\end{center}
\end{figure}

  Two obvious forms of VBC order to consider in this context are the
  plaquette valence-bond crystalline (PVBC) and crossed-dimer valence-bond
  crystalline (CDVBC) forms illustrated in Figs.\
  \ref{VBC_patterns}(a) and (b) respectively.
For both cases we simply consider the response of the system when a
corresponding field operator, $F=\delta\; \hat{O}$, is added as a
small perturbation to the original Hamilton of Eq.\ (\ref{H}),
with $\delta$ an infinitesimally small $c-$number.\cite{darradi08}
The particular operators $\hat{O}_{p}$ and $\hat{O}_{d}$,
corresponding respectively to PVBC and CDVBC ordering, are illustrated
graphically in Figs.\ \ref{VBC_patterns}(a) and (b) and are also defined
explicitly in the caption.

In both cases we calculate the perturbed GS energy per spin, $e(\delta) \equiv E(\delta)/N$, for the
perturbed Hamiltonian $H + F$, at various LSUB$m$ levels of
approximation.  We use the N\'{e}el and DN states as CCM model
states since we are especially interested in the phase intermediate between them in the spin-1/2 phase diagram.  We then calculate the corresponding susceptibility, 
\begin{equation}
\chi \equiv -\left. (\partial^2{e(\delta)})/(\partial {\delta}^2)   \label{Eq_X}
\right|_{\delta=0}\,,
\end{equation}
and use it to find points or regions where the phase corresponding to
the particular CCM model state used becomes unstable against the
specified form of VBC order, namely when its extrapolated inverse
susceptibility, $\chi^{-1}$, goes to zero.

Clearly our CCM LSUB$m$ results for any susceptibility still need to be extrapolated to the LSUB$\infty$ limit.  The most straightforward way to do so \cite{Li:2013_chevron} is clearly to extrapolate first our LSUB$m$ results for the perturbed energy using an unbiased scheme such as in Eq.\ (\ref{M_extrapo_nu}),
\begin{equation}
e^{(m)}(\delta) =
e_{0}(\delta)+e_{1}(\delta)m^{-\nu}\,,   \label{Extrapo_dBonds}
\end{equation}
with the exponent $\nu$ a fitting parameter, along with
$e_{0}(\delta)$ and $e_{1}(\delta)$.  Generally, as is to be expected
from our standard LSUB$m$ energy extrapolation scheme of Eq. 
(\ref{E_extrapo}), the fitted value of $\nu$ is close to 2
for most values of the frustration parameter $\alpha$
pertaining to the particular CCM model state used, except very near
(or inside) any critical regime, where it can deviate sharply from the
value 2, as discussed in more detail below.

\begin{figure*}[!t]
\begin{center}
\mbox{
\subfigure[]{\scalebox{0.31}{\includegraphics[angle=270]{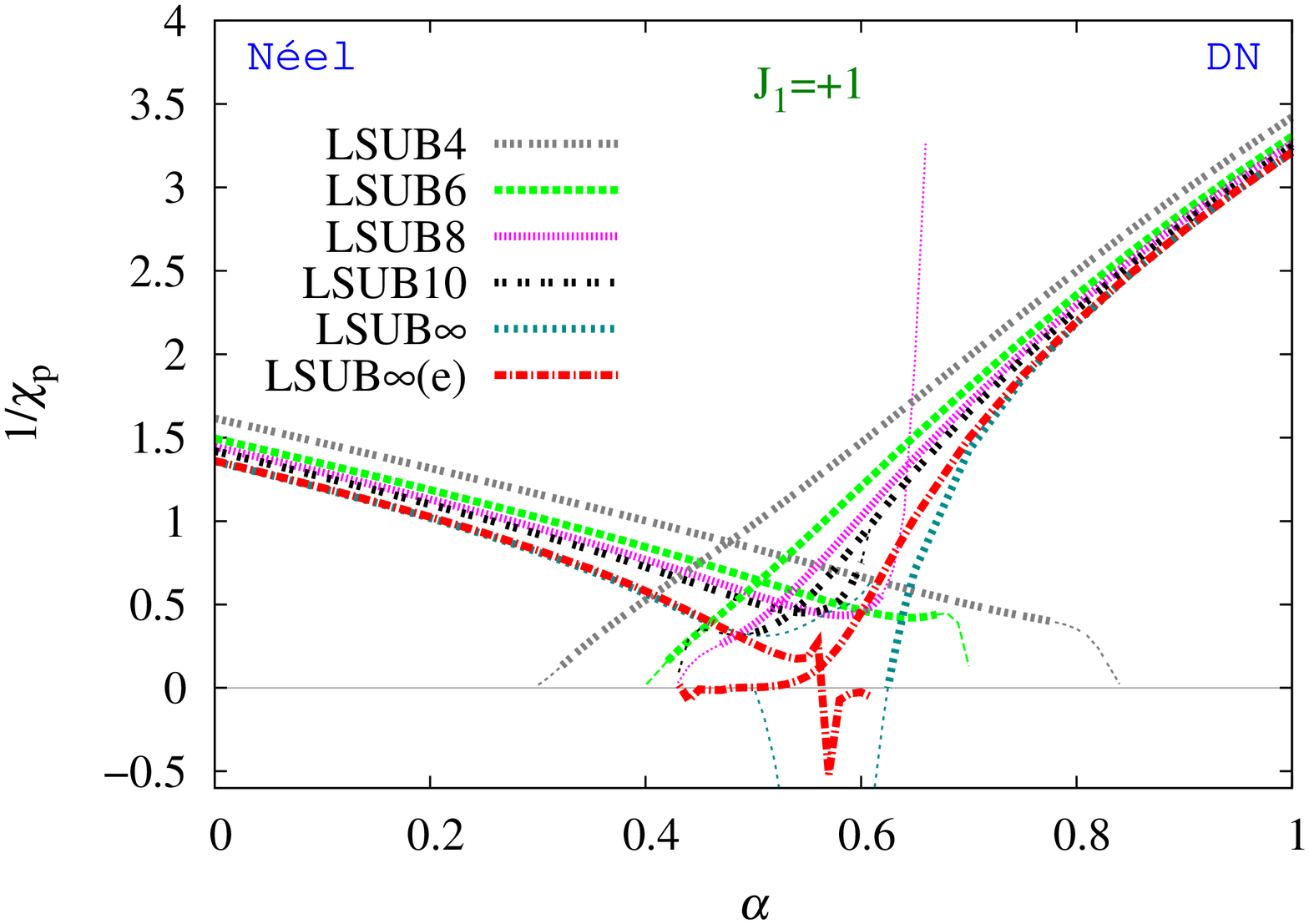}}} 
\subfigure[]{\scalebox{0.31}{\includegraphics[angle=270]{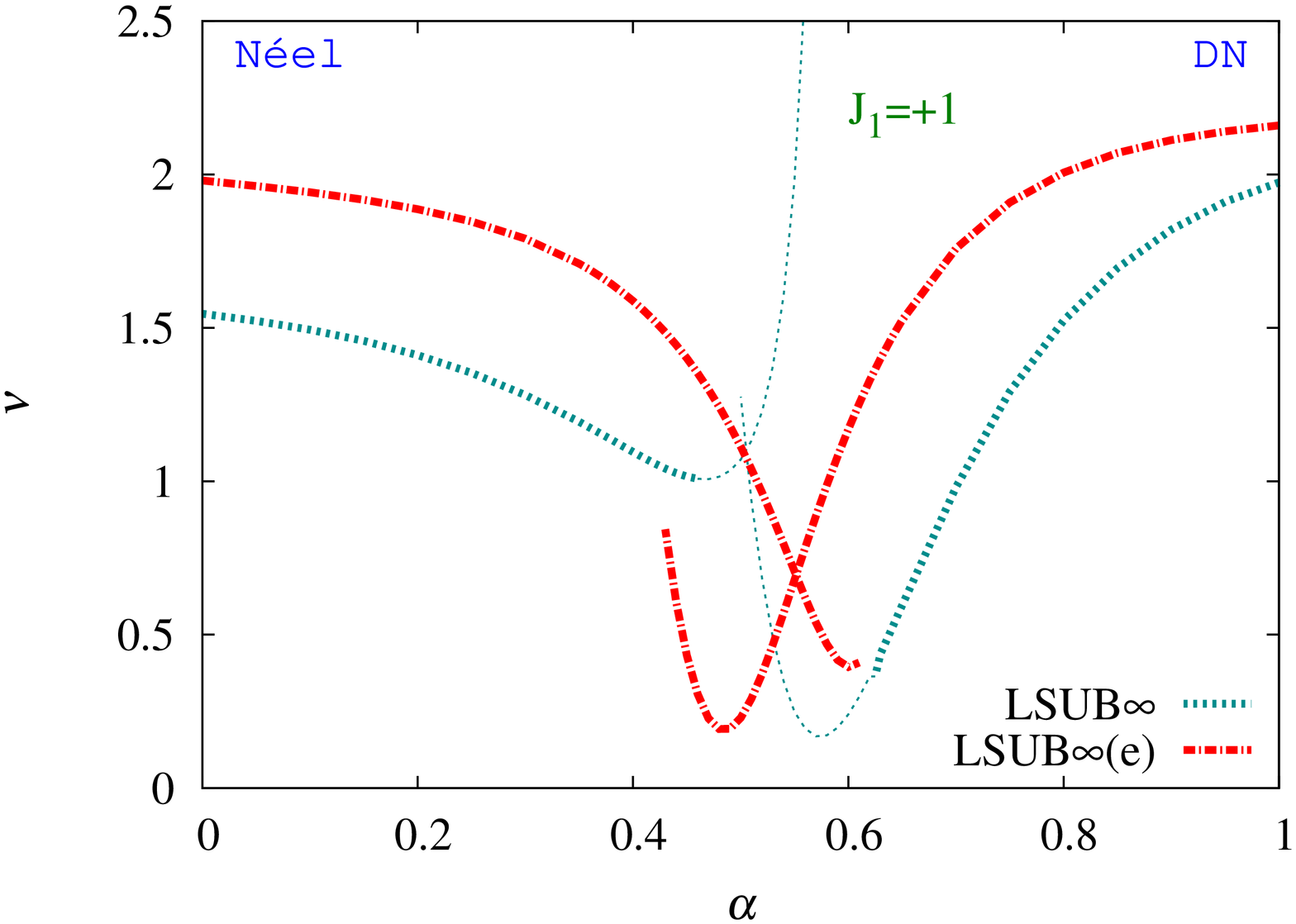}}} 
}
\caption{(Color online) (a) CCM results for the inverse plaquette
  susceptibility, $1/\chi_{{\rm p}}$, appropriate to the PVBC ordering
  of Fig.\ \ref{VBC_patterns}(a), as a function of the frustration
  parameter, $\alpha \equiv J_{2}/J_{1}$, for the spin-1/2
  $J_{1}$--$J_{2}$ model (with $J_{1}=+1$) on a cross-striped square
  lattice, using both the N\'{e}el state (left curves) and DN state
  (right curves) as the CCM model state.  For both model states we
  show the LSUB$m$ results with $m=\{4,6,8,10\}$, plus the
  corresponding extrapolated results LSUB$\infty$ and LSUB$\infty$(e)
  using this data set in the extrapolation schemes of Eqs.\
  (\ref{Extrapo_inv-chi-2}) and (\ref{Extrapo_dBonds}), respectively.
  Those portions of the curves with thinner lines indicate the
  respective unphysical regions where $M<0$ (and see text for
  details).  Note that the LSUB$\infty$ curves are shown with thinner
  lines over the unphysical ranges, since the corresponding results for
  $1/\chi_{p}$ clearly behave non-monotonically there.  By contrast,
  the LSUB$\infty$(e) curves are shown over the entire region where
  all of the LSUB$m$ approximations have real solutions (i.e., out
  to the respective termination points of the highest, LSUB10,
  solution), since the results for the perturbed energies used in this
  extrapolation behave monotonically almost everywhere the solutions exist.
  (b) The respective fitted values of the exponent $\nu$ in the
  extrapolation schemes of Eqs.\ (\ref{Extrapo_inv-chi-2}) and
  (\ref{Extrapo_dBonds}), for the same fits as shown in Fig.\
  \ref{X_plaqXcpty}(a) to the left.}
\label{X_plaqXcpty}
\end{center}
\end{figure*}    

We have also found previously\cite{DJJF:2011_honeycomb,Li:2013_chevron} that
our LSUB$m$ values $\chi(m)$ may themselves very accurately be directly
extrapolated to the $m \rightarrow \infty$ limit using the same scheme,
\begin{equation}
\chi(m) = d_{0}+d_{1}m^{-2}+d_{2}m^{-4}\,,  \label{Extrapo_X_asE}
\end{equation}
as for the GS energy itself in Eq.\ (\ref{E_extrapo}).  A
corresponding direct extrapolation of the more relevant quantity, the
inverse susceptibility,
\begin{equation}
\chi^{-1}(m) = x_{0}+x_{1}m^{-2}+x_{2}m^{-4}\,,             \label{Extrapo_inv-chi-1}
\end{equation}
has also been found\cite{DJJF:2011_honeycomb,Li:2013_chevron} to give
consistently good results, which agree well with those obtained from
Eq.\ (\ref{Extrapo_X_asE}), although again with the exception of regions
where $\chi^{-1}$ becomes very small or zero.  Since we are here
interested precisely in such regions over an extended range of
values of the frustration parameters, $\alpha^{c}_{1a} \lesssim \alpha \lesssim \alpha^{c}_{1b}$, we may also use an unbiased extrapolation scheme of the form of Eq.\ (\ref{M_extrapo_nu}),
\begin{equation}
\chi^{-1}(m) = y_{0}+y_{1}m^{-\nu}\,,            \label{Extrapo_inv-chi-2}
\end{equation}
in such a case,\cite{DJJF:2011_honeycomb,Li:2013_chevron} where
$y_{0}$, $y_{1}$, and $\nu$ are all treated as fitting parameters.

In Fig.\ \ref{X_plaqXcpty}(a) we present our results for the inverse
plaquette susceptibility, $\chi^{-1}_{{\rm p}}$, pertaining to the
PVBC ordering illustrated graphically in Fig.\ \ref{VBC_patterns}(a).
We note that since the definition of $\chi$ is invariant with respect
to the sign of the perturbation parameter $\delta$ in the case of PVBC
order, its graphical definition in Fig.\ \ref{VBC_patterns}(a) is
invariant with respect to interchange of the strengthened (thick, red)
and weakened (thin, blue) $J_{1}$ bonds.  We show explicitly in Fig.\
\ref{X_plaqXcpty}(a) our LSUB$m$ results based on both the N\'{e}el
and DN states as CCM model states, with $m=\{4,6,8,10\}$, together
with two extrapolated results, LSUB$\infty$ and LSUB$\infty$(e), based
on Eqs.\ (\ref{Extrapo_inv-chi-2}) and (\ref{Extrapo_dBonds})
respectively, and in each case using the respective data sets
$m=\{4,6,8,10\}$ to perform the fits.  What is especially noteworthy
in the first place is how very close are the two different
extrapolations for both the N\'{e}el and DN states as model states,
except precisely in the region $\alpha^{c}_{1a} \lesssim \alpha
\lesssim \alpha^{c}_{1b}$ where they have different forms.  However,
even in this most interesting region, the physical picture that
emerges is a rather consistent one.

Thus, from the raw LSUB$m$ results themselves, we see clearly that
both the N\'{e}el-ordered and DN-ordered states become highly
susceptible to PVBC ordering around the same points at which their
respective magnetic order parameters, $M$, approach zero, as in Fig.\
\ref{M}(a).  On the N\'{e}el side, although the LSUB$\infty$ result
for $\chi^{-1}_{p}$ based on Eq.\ (\ref{Extrapo_inv-chi-2}) does
not become exactly zero it does become very small around our previous
estimates for $\alpha^{c}_{1a}$, and the corresponding LSUB$\infty$(e)
result shows a clear minimum, with an even smaller value of
$\chi^{-1}_{p}$, at a slightly larger value of $\alpha$, just before the
extrapolation becomes unstable, in the region where the corresponding
solutions are unphysical since (some of) the LSUB$m$ solutions have a
value there of $M<0$, as seen from Fig.\ \ref{M}(a).

By contrast, the extrapolated results based on the DN state have
markedly different character.  Thus, the LSUB$\infty$ extrapolated
result for $1/\chi_{p}$ based on Eq.\ (\ref{Extrapo_inv-chi-2})
goes to zero at a value $\alpha=0.62(2)$, where the error bar is an
estimate from using different LSUB$m$ data sets, as discussed
previously.  A close inspection of Fig.\ \ref{X_plaqXcpty}(a)
reveals, however, that this LSUB$\infty$ result for $1/\chi_{p}$ based
on the extrapolation scheme of Eq.\ (\ref{Extrapo_inv-chi-2}) then
becomes zero again as $\alpha$ is decreased further, at a value
$\alpha \approx 0.50$.  These two values are completely consistent
with our previous estimates of the two QCPs marking the range,
$\alpha^{c}_{1a} \leq \alpha \leq \alpha^{c}_{1b}$, of the
intermediate phase, namely, $\alpha^{c}_{1b}=0.62(1)$ and
$\alpha^{c}_{1a}=0.46(1)$.  Even more revealing perhaps is the
estimate LSUB$\infty$(e) for $1/\chi_{p}$, shown in Fig.\ \ref{X_plaqXcpty}(a), which is based on the most direct extrapolation
scheme of Eq.\ (\ref{Extrapo_dBonds}).  Here we observe very
clearly that $1/\chi_{p}$ is zero (or very close to zero within the
small error bars of the extrapolation) over a range $\alpha^{c}_{1a}
\lesssim \alpha \lesssim \alpha^{c}_{1b}$.  Thus, all of the evidence
from our results from $\chi_{p}$ is compatible with the interpretation
that the quantum phase intermediate between those with N\'{e}el and DN
order has PVBC order.

In Fig.\ \ref{X_plaqXcpty}(b) we show the values of the exponent
$\nu$ that we obtain from our LSUB$\infty$ and LSUB$\infty$(e) fits to
the LSUB$m$ extrapolation schemes of Eqs.\
(\ref{Extrapo_inv-chi-2}) and (\ref{Extrapo_dBonds}) respectively.
The values shown are based on fitting to the LSUB$m$ data set
$m=\{4,6,8,10\}$ for both the N\'{e}el and DN solutions.  However, the
fitted values are themselves again remarkably robust with respect to
the choice of data set.  Figure \ref{X_plaqXcpty}(b) shows that when
$\chi_{p}$ is calculated using Eq.\ (\ref{Extrapo_dBonds}), the
fitted value of the exponent $\nu$ is very close to the expected value
2, as in our standard GS energy extrapolation scheme of Eq.\ 
(\ref{E_extrapo}), except for values of $\alpha$ in the approximate
range $\alpha^{c}_{1a} \lesssim \alpha \lesssim \alpha^{c}_{1b}$,
where it drops sharply.  In practice the second derivative of
$e(\delta)$ required in Eq.\ (\ref{Eq_X}) is calculated
numerically using values $\delta=0,\pm d$, typically with $d=0.001$.
The corresponding value of $\nu$ in Eq.\ (\ref{Extrapo_dBonds})
that are plotted in Fig.\ \ref{X_plaqXcpty}(b) are then essentially
identical for these three values of $\delta$ that we use.  On the
other hand, when $1/\chi_{p}$ is calculated using Eq.\ 
(\ref{Extrapo_inv-chi-2}) the fitted value for the exponent $\nu$ is
close to 1.5 on the N\'{e}el side and 2 on the DN side, again except
for values of $\alpha$ in the critical range $\alpha^{c}_{1a} \lesssim
\alpha < \alpha^{c}_{1b}$, where they similarly deviate sharply.  It
is interesting to note that a value $\nu \approx 1.5$ has also been
observed previously when using Eq.\ (\ref{Extrapo_inv-chi-2}) for
the extrapolations of $1/\chi_{p}$ for the N\'{e}el state in the
related $J_{1}$--$J_{2}$ models on the checkerboard
\cite{Bishop:2012_checkerboard} and chevron-square\cite
{Li:2013_chevron} lattices.

Although our results for $\chi_{p}$ provide very strong evidence for a
PVBC-ordered GS phase intermediate between the N\'{e}el and DN GS
phases for the quantum spin-1/2 model, we have also performed similar
CCM calculations based on the N\'{e}el and DN states as model states
for the corresponding crossed-dimer susceptibility, $\chi_{d}$,
pertaining to the CDVBC order illustrated graphically in Fig.\
\ref{VBC_patterns}(b).  The results for $\chi_{d}$ are
qualitatively quite different to those for $\chi_{p}$.  Thus, in the
case of $\chi_{d}$ the extrapolated results show no indication at all
of being zero (or unphysically negative) over any finite range.
Instead, the DN and N\'{e}el results, respectively, for $1/\chi_{d}$
become zero (or very closely approach zero) only at single points,
which are themselves completely compatible with our prior estimates for
$\alpha^{c}_{1a}$ and $\alpha^{c}_{1b}$.  Thus, while the results for
$1/\chi_{d}$ corroborate our previous estimates for these two QCPs,
they provide no evidence at all for any CDVBC-ordered phase, since
$1/\chi_{d}$ does not vanish over any finite range of values of
$\alpha$.  The fact that $1/\chi_{d}$ vanishes at specific points,
namely $\alpha^{c}_{1a}$ and $\alpha^{c}_{1b}$, simply reinforces
these as being QCPs, since at any QCP one expects the system to become
infinitely susceptible to {\it all} forms of ordering that are
compatible with the symmetries of the physical and model states.

We now summarize our results in Sec.\ \ref{summary_sec}.

\section{SUMMARY}
\label{summary_sec}
We have investigated the complete $T=0$ GS phase diagram of an $s=\frac{1}{2}$ $J_{1}$--$J_{2}$ Heisenberg model on a cross-striped square lattice, for all values of $J_{1}$ and $J_{2}$, both positive and negative.   
\begin{figure*}
\vskip1.4cm
\begin{center}
\includegraphics[width=7cm,angle=90]{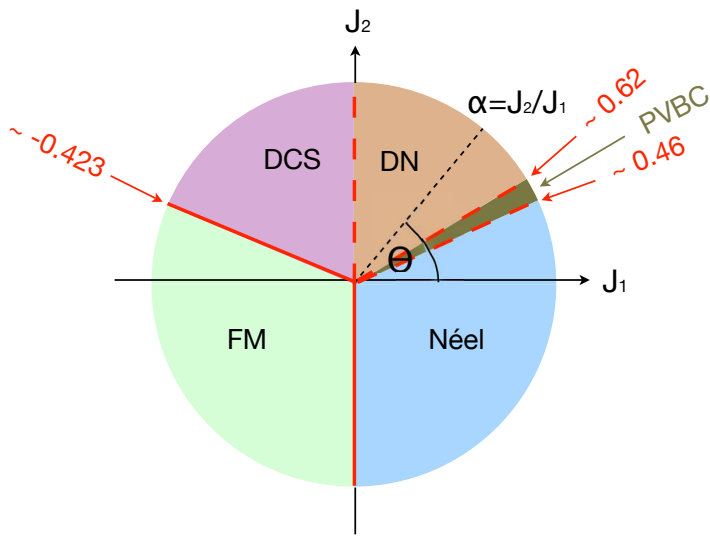}
\end{center}
\caption{(Color online) The ground-state phase diagram of the spin-1/2
  $J_{1}$--$J_{2}$ model on a cross-striped square lattice (with
  $\alpha \equiv \tan \theta \equiv J_{2}/J_{1}$), showing the three
  antiferromegnetic phases, namely N\'{e}el, double N\'{e}el (DN) and
  double columnar striped (DCS), illustrated in Figs.\ \ref{model_bonds}(a)--(c) respectively; the plaquette valence-bond
  crystalline (PVBC) phase illustrated in Fig.\ \ref{VBC_patterns}(a);
  and the ferromagnetic (FM) phase.  The three transitions at
  $\alpha^{c}_{1a}=0.46(1)$ and $\alpha^{c}_{1b}=0.62(1)$, which mark
  the boundaries of the PVBC phase, and at
  $\theta^{c}_{2}=\frac{1}{2}\pi$ between the DN and DCS phases, all
  with $J_{2}>0$, each shown by a broken line, all appear to be
  continuous ones; while the two transitions at
  $\alpha^{c}_{3}=-0.423(1)$ (with $J_{2}>0$) and at
  $\theta^{c}_{4}=\frac{3}{2}\pi$, which mark the boundaries of the FM
  phase, both shown by solid lines, are first-order ones.}
\label{phase-diagram}
\end{figure*}
The classical ($s \rightarrow \infty$) version of the model has four GS
phases, as illustrated in Fig.\ \ref{model_bonds}(d), with each of
the corresponding phase transitions of first-order type.  In the first
quadrant of the phase diagram (where $J_{1} \geq 0$ and $J_{2} \geq
0$) the model interpolates continuously between a 2D HAF on the square
lattice (when $J_{2}/J_{1} \equiv \alpha = 0$) and uncoupled 1D HAF
chains (as $\alpha \rightarrow \infty$).  For the spin-1/2 quantum
model we have found a GS phase diagram with five phases, with our main
findings summarized in Fig.\ \ref{phase-diagram}.

One of our main conclusions is that the classical direct first-order
transition between the AFM N\'{e}el and DN phases at $\alpha^{{\rm
    cl}}_{1}=0.5$ (with $J_{1}>0$) is split into two transitions in
the $s=\frac{1}{2}$ case, with QCPs at $\alpha^{c}_{1a}=0.46(1)$ and
$\alpha^{c}_{1b}=0.62(1)$, and an intermediate quantum phase with PVBC
ordering.  From the shape of the order-parameter curves in Fig.\
\ref{M}(a) it seems probable that both transitions are continuous,
since a first-order transition is usually signalled by a much steeper
(or discontinuous) fall to zero.\cite{DJJF:2011_honeycomb}  The shape
of the corresponding curves for $1/\chi_{p}$ in Fig.\
\ref{X_plaqXcpty}(a) also corroborates that the transitions are
continuous, since first-order ones also usually show a similar steep
(or discontinuous) drop to zero.\cite{DJJF:2011_honeycomb}  Since the
three phases, N\'{e}el, DN and PVBC, break different symmetries our
results thus favour the deconfinement scenario for both transitions at
$\alpha^{c}_{1a}$ and $\alpha^{c}_{1b}$.  Nevertheless, we should mention that
generic arguments have been given\cite{Kuklov:2008} that phase transitions
in models with SU(2)-symmetric deconfined critical points should be of
first-order type.  However, these arguments are based on effective
field theories, while our own calculations are based directly on the
lattice model itself.  Although we can never entirely exclude the
possibility of the transitions at $\alpha^{c}_{1a}$ and
$\alpha^{c}_{1b}$ being sufficiently weak first-order ones, our
evidence points more clearly to them being of second-order
(deconfined, continuous) type.  While our most accurate determination
of both $\alpha^{c}_{1a}$ and $\alpha^{c}_{1b}$ comes from the
magnetic order parameter results shown in Figs.\ \ref{M}(a) and
\ref{M_Extrapolation}(a) as the respective points where N\'{e}el order
and DN order melt, our results for $1/\chi_{p}$ shown in Fig.\
\ref{X_plaqXcpty}(a) strongly corroborate that these are the same
points where PVBC order turns on.  Naturally, we cannot entirely
exclude, from our results as shown in Fig.\ \ref{X_plaqXcpty}(a), the
possibility of a very narrow regime within the range $\alpha^{c}_{1a}
< \alpha < \alpha^{c}_{1b}$ where yet another phase with a different
form of ordering exists.

We appreciate that the evidence presented here for the transitions at
$\alpha^{c}_{1a}$ and $\alpha^{c}_{1b}$ being of the continuous
(deconfined) type is relatively weak and rather far from being
conclusive.  Nevertheless, we believe that the analysis is certainly
sufficiently suggestive to justify further work to clarify the nature
of these transitions, for example, by the calculation of critical
exponents or by finding a positive signal of the emergent [U(1)]
symmetry.  Such calculations, however, have scarcely ever hitherto been
attempted within the CCM framework itself, and are certainly beyond
the scope of the present study in any case.

We have found, as expected, that the third QCP at
$\theta^{c}_{3}=\frac{1}{2}\pi$ between the DN and DCS phases,
coincides with the corresponding classical transition, $\theta^{{\rm
    cl}}_{3}=\frac{1}{2}\pi$.  However, we have found very strong
evidence, from both the GS energy results shown in Fig.\ \ref{E}(b)
and the magnetic order parameter results shown in Figs.\ \ref{M}(b)
and \ref{M_Extrapolation}(b), that the quantum transition for the
$s=\frac{1}{2}$ model is again a continuous second-order one (and
hence, again, presumably a deconfined transition), by contrast with
the first-order nature of this transition in the classical ($s
\rightarrow \infty$) model.

On the other hand, the remaining two transitions that mark the boundaries of
the FM phase, are clearly first order in both the $s=\frac{1}{2}$ and
$s \rightarrow \infty$ cases.  However, we have found that quantum
fluctuations act to stabilize the collinear AFM order of the DCS phase
to higher values of the corresponding frustration (i.e., here, to
larger values of $\alpha^{-1}$, since the $J_{1}$ bonds now act to
frustrate the AFM order of the $J_{2}$ chains) than in the classical
case.  Thus, we found that the classical transition at $\alpha^{{\rm
    cl}}_{3}=-0.5$ (with $J_{1}>0$) shifts by quantum fluctuations in
the $s=\frac{1}{2}$ case to $\alpha^{c}_{3}=-0.423(1)$, where our best
estimate for $\alpha^{c}_{3}$ now comes from the energy crossing point
shown in Fig.\ \ref{E}(c).  Such stabilization by quantum
fluctuations of collinear AFM order at the expense of FM order in
frustrated regions has been observed elsewhere, for example, in both
the FM version of the full (undepleted) spin-1/2
$J_{1}$--$J_{2}$ model on the square lattice\cite{richter10:J1J2mod_FM} and in a related model on the honeycomb
lattice.\cite{Bishop:2012_honey_phase}

Lastly, in the unfrustrated regime where $J_{2}<0$, the final QCP
between the FM and the N\'{e}el phases has been found to occur at
$\theta^{c}_{4}=\frac{3}{2}\pi$, at precisely the same place as the
corresponding classical transition, $\theta^{{\rm
    cl}}_{4}=\frac{3}{2}\pi$, fully as expected.

As final point, it may be worthwhile to comment on the limitations of
the present CCM formalism in this context.  While there exists a large
amount of strong evidence that the method can very accurately capture
the properties and phase boundaries of (magnetically) ordered states
of highly frustrated quantum magnets, the available evidence for its
ability to capture phases that are not adiabatically connected to a
chosen reference state is mixed.  On the one hand there is
considerable evidence, including from the present study, that the CCM
can well describe the phase boundaries of such states as those without
magnetic order but with various forms of VBC order, even when using a
reference state with magnetic order that is not itself the stable GS
phase in the region (or, indeed, anywhere).  On the other hand, and in
common with many other methods, the CCM does not easily detect
directly such disordered phases as spin-liquid phases, e.g., of the
topological spin-liquid type or the sliding Luttinger liquid (SLL)
type mentioned in Sec.\ \ref{model_sec}.  What the CCM can perhaps
most easily provide in such circumstances is strong evidence for a
region in the ($T=0$) GS phase diagram of a phase of a type for which
one may then test by other means.  In other words, in such
circumstances, it is better suited to exclude possibilities and/or to
provide signals for the existence of some (as yet unknown) phase.  For
example, for the present model it is conceivable that an SLL might,
{\it a priori}, exist at high enough (but still finite) values of
$\alpha$.  However, as we have seen, no indications emerge from the
present analysis that would lend credence to, or would justify a
search for, such an SLL phase as a stable GS phase.

In conclusion, we have seen that the $J_{1}$--$J_{2}$ Heisenberg model
on the cross-striped square lattice provides a challenging model with
a rich GS phase diagram in the extreme $s=\frac{1}{2}$ quantum case,
with several features that differ markedly from its classical ($s
\rightarrow \infty$) counterpart.  The application of other
theoretical techniques to the model would hence surely be of interest,
in order to confirm our results.  It might also be interesting to
examine the $s=1$ version of the model for further differences.

\section*{ACKNOWLEDGMENTS}
We thank the University of Minnesota Supercomputing Institute for the
grant of supercomputing facilities for this research.

\end{document}